\documentclass[11pt, a4paper]{article}

\usepackage{amsmath,amsfonts,bigints} 
\usepackage{fullpage}
\usepackage{my_package}
\usepackage{color}
\usepackage{tikz}
\usepackage{tkz-euclide}
\usepackage{url}
\usepackage{authblk}

\title{A microstructural model of transversely isotropic, fibre-reinforced hydrogels}

\author[1]{Matthew G.~Hennessy\footnote{Equal contributions. Emails: {\tt matthew.hennessy@bristol.ac.uk}, {\tt tom.shearer@manchester.ac.uk}}}
\author[2]{Tom Shearer$^*$}
\author[3]{Axel C.~Moore}
\affil[1]{School of Engineering Mathematics and Technology, University of Bristol, Bristol, UK}
\affil[2]{Department of Mathematics, University of Manchester, Manchester, UK}
\affil[3]{Department of Biomedical Engineering, Carnegie Mellon University, Pittsburgh, PA, USA}
\date{}

\begin{document}

\maketitle

\begin{abstract}
Fibre-reinforced hydrogels are promising
materials for biomedical applications due to
their strength, toughness, and tunability. 
However, it remains unclear how to design
fibre-reinforced hydrogels for use in specific applications due to the lack of a flexible modelling framework that can
predict and hence optimise their behaviour.
In this paper,
we present a microstructural model for transversely isotropic
fibre-reinforced hydrogels that
captures the specific geometry of the fibre
network.  The model also accounts for slack in the initial
fibre network that is gradually removed upon deformation. The 
mechanical model for the fibre network 
is coupled to a nonlinear poroelastic model for the
hydrogel matrix that accounts for osmotic
stress.  By comparing the model predictions
to data from unconfined compression experiments, we show that the model can capture J-shaped stress-strain
curves and time-dependent creep responses.
We showcase how the model can be used to 
guide the design of materials for artificial
cartilage by exploring how to maximise
interstitial fluid pressure.  We find that 
fluid pressurisation can be increased
by using stiffer fibres, removing slack
from the fibre network, and reducing the
Young's modulus of the hydrogel matrix.
Finally, a high-level and open-source Python package has been developed for simulating unconfined
compression experiments using the model.

\end{abstract}

\newpage
\begin{table}[ht]
\centering
\caption{Nomenclature.}
\label{tab:nomenclature}
\footnotesize
\begin{tabular}{|c|l|} \hline
Parameter & Definition\\
\hline
UC & Unhydrated (nominal) configuration\\
HC & Hydrated (referential)  configuration\\
CC & Current (deformed) configuration\\
$A$ & Area of circular faces of cylindrical sample in HC \\
$\tens{C}=\tens{F}^T\tens{F}$ & Right Cauchy-Green tensor of post-hydration deformation (from HC to CC)\\
$E_f$ & Fibre Young's modulus\\
$E_m$ & Matrix Young's modulus\\
$\tens{F}_t$ & Total deformation gradient (from UC to CC)\\
$\tens{F}_h$ & Deformation gradient associated with hydration (from UC to HC)\\
$\tens{F}$ & Deformation gradient of post-hydration deformation (from HC to CC)\\
$F$ & Force\\
$f(\lambda_c)$ & Probability density function of fibre segment waviness distribution\\
$h(t)$ & Height of cylindrical sample in CC over time\\
$\tens{I}$ & Identity tensor\\
$I_1$, $I_2$, $I_3$ & Isotropic strain invariants\\
$I_4$, $I_5$ & Anisotropic strain invariants\\
$J_t$ & Determinant of $\tens{F}_t$\\
$J_h$ & Determinant of $\tens{F}_h$\\
$J$ & Determinant of $\tens{F}$\\
$k$ & Permeability of material in CC\\
$k_0$ & Permeability of material in HC\\
$N$ & Number of fibre segments in the network\\
$p$ & Fluid pressure (force per unit area of CC)\\
$\bar{p}$ & Radially averaged fluid pressure\\
$\vec{Q}_0$ & Volume flux of water per unit area of UC\\
$\vec{Q}_w$ & Volume flux of water per unit area of HC\\
$r$ & Radial variable in HC\\
$r_c(t)$ & Radius of cylindrical sample in CC over time\\
$R$ & Initial radius of cylindrical sample in HC\\
$R_g$ & Universal gas constant\\
$S$ & Fibre segment undeformed length\\
$s$ & Fibre segment deformed length\\
$\tens{S}_0$ & First Piola--Kirchhoff (PK)
or nominal stress tensor (forces in CC per unit area of UC)\\
$\tens{S}$ & Referential stress tensor (forces in CC per unit area of HC)\\
$S_r$, $S_\theta$, $S_z$ & Radial, azimuthal and longitudinal components of $\tens{S}$\\
$S_i$ & $i^\text{th}$ diagonal component of $\tens{S}$\\
$\tens{S}^e_0$ & Nominal elastic stress (forces in CC per unit area of UC)\\
$\tens{S}^e$ & Referential elastic stress (forces in CC per unit area of HC)\\
$S^e_r$, $S^e_\theta$, $S^e_z$ & Radial, azimuthal and longitudinal components of $\tens{S}^e$\\
$S_i^e$ & $i^\text{th}$ diagonal component of $\tens{S}^e$\\
$T$ & Temperature\\
$t$ & Time \\
$U_f^{(i)}$ & Strain energy of $i^\text{th}$ fibre segment\\
$\bar{U}_f$ & Mean strain energy of all fibre segments\\

\hline
\end{tabular}
\end{table}

\begin{table}
\centering
\caption{Nomenclature -- continued.}
\label{tab:nomenclature2}
\footnotesize
\begin{tabular}{|c|l|} \hline
Parameter & Definition\\
\hline
$U_f^\text{tot}$ & Total strain energy of all fibre segments in network\\
$\mathcal{U}_{[a,b]}$ & Continuous uniform probability distribution over the interval $[a,b]$ \\
$\vec{u}_t$ & Displacement relative to UC\\
$\vec{u}$ & Displacement relative to HC\\
$u_r$ & Radial displacement relative to HC\\
$V_f^{(i)}$ & Volume of $i^\text{th}$ fibre segment\\
$\bar{V}_f$ & Mean volume of fibre segments\\
$V_f^\text{tot}$ & Total volume of all fibre segments in network\\
$V_s$ & Volume of solid in the UC, HC and CC (due to assumed incompressibility)\\
$V_w$ & Volume of water in the HC\\
$V_h$ & Total volume of the material in the HC\\
$W_f^{(i)}$ & Strain energy density of an individual fibre segment\\
$\tilde{W}_f $ & Strain energy density of fibres averaged over distribution of critical stretches\\
$W_f$ & Strain energy density of fibres averaged over distribution of angles\\
$W_m$ & Strain energy density of matrix phase (per unit volume of UC)\\
$W_e$ & Strain energy density of entire elastic phase (per unit volume of UC)\\
$W_\text{mix}$ & Energy density of mixing of water molecules (per unit volume of UC)\\
$W$ & Energy density of entire material (per unit volume of UC)\\
$\alpha_\parallel$, $\alpha_\perp$ & Stretches from UC to HC parallel and perpendicular to plane of isotropy\\
$\beta_r$, $\beta_\theta$, $\beta_z$ & Stretches from HC to CC in radial, azimuthal and longitudinal directions\\
$\gamma$, $M$ & Fitting parameters in Holmes-Mow permeability equation\\
$\Delta X$, $\Delta Y$ & Initial side lengths of triangle formed by fibre segment in UC \\
$\Delta x$, $\Delta y$ & Current side lengths of triangle formed by fibre segment in CC \\
$\Theta$ & Initial angle fibre segment makes with $X$-axis\\
$\theta$ & Deformed angle fibre segment makes with $X$-axis\\
$\lambda=s/S$ & Fibre segment stretch\\
$\lambda_c$ & Critical stretch at which wavy fibre segment tautens\\
$\lambda_m$ & Maximum recruitment stretch needed to engage all fibres in the network\\
$\lambda_1$, $\lambda_2$, $\lambda_3$ & Total stretches (from UC to CC) in $X$-, $Y$-, $Z$- directions\\
$\lambda_\parallel$, $\lambda_\perp$ & Total stretches parallel and perpendicular to plane of isotropy\\
$\lambda_r$, $\lambda_\theta$, $\lambda_z$ & Total stretches in radial, azimuthal and longitudinal directions\\
$\mu_w$ & Chemical potential of water\\
$\nu_m$ & Matrix Poisson's ratio\\
$\Pi$ & Osmotic stress (force per unit area of CC)\\
$\pi$ & Pi ($\approx 3.14159$, not related to $\Pi$)\\
$\rho_m$, $\rho_f$, $\rho_w$ & mass density of the matrix, fibres, and water, respectively \\
$\tilde{\sigma}_f$ & Cauchy stress of fibres averaged over distribution of critical stretches\\
$\tens{\sigma}$ & Total Cauchy stress (forces in CC per unit area of CC)\\
$\tens{\sigma}^e$ & Elastic Cauchy stress (forces in CC per unit area of CC)\\
$\sigma^e_\parallel$, $\sigma^e_\perp$ & Elastic Cauchy stresses parallel and perpendicular to plane of isotropy\\
$\Phi_f$ & Volume fraction of fibres in UC, HC and CC per unit volume of UC\\
$\Phi_w$ & Volume fraction of water in CC per unit volume of UC\\
$\phi_0$ & Volume fraction of water in HC per unit volume of HC (initial porosity) \\
$\phi_w$ & Volume fraction of water in CC per unit volume of HC\\ 
$\phi$ & Volume fraction of water in CC per unit volume of CC (current porosity)\\ 
$\chi$ & Flory interaction parameter\\
$\Omega_w$ & Molar volume of water\\
$\nabla$ & Gradient with respect to spatial coordinates in HC \\
$\nabla_0$ & Gradient with respect to spatial coordinates in UC \\
$\hat{\cdot}$ & Hats used to distinguish 3D quantities from 2D quantities in Sections \ref{micro}-\ref{combining}\\
\hline
\end{tabular}
\end{table}

\newpage

\section{Introduction}


Hydrogels consist of highly hydrated networks of polymer chains, which are used in diverse applications ranging from contact lenses \cite{ishihara2023biomimetic} to soft robotics \cite{chen2023bioinspired}. They are valued as biomaterials due to their biocompatibility \cite{nasra2023functional}, biodegradability \cite{zhang2021biodegradable}, and tuneable physical properties \cite{smith2021alginate}, and are being increasingly used in regenerative medicine \cite{revete2022advancements}.  However,
hydrogels suffer from low strength and
toughness owing to their high water content,
and they are often unable to satisfy the
mechanical demands of biomedical applications~\cite{agrawal2013}.  Various
approaches have been proposed to strengthen
and toughen hydrogels~\cite{fuchs2020},
which include increasing the uniformity of
the polymer network~\cite{sakai2008},
creating multifunctional crosslinks 
using nanostructure additives~\cite{haraguchi2002}, and
adding a second, interpenetrating polymer
network~\cite{dragan2014}. 
Hydrogels that are reinforced by a fibre network have shown
particular promise in biomedicine~\cite{beckett2020} as the
strength and toughness of the fibre network removes the
need to use highly crosslinked hydrogels, which can lead
to unfavourable environments for cells.  In addition, 
fibre-reinforced hydrogels offer a great deal of tunability,
as the fibre stiffness~\cite{tan2023}, 
volume fraction~\cite{visser2015, jordan2017, moore2023}, and 
alignment~\cite{bas2015, castilho2019}
can all be varied
to produce composite materials with optimal mechanical
and poroelastic responses.  


Despite the advantages of fibre-reinforced hydrogels, there
remain challenges in engineering these materials for use 
in specific applications.  In particular, there is a lack of
understanding of how the fibre 
properties and the geometry of the fibre network
should be varied in order to produce tailored macroscropic 
responses~\cite{beckett2024}.  
Mathematical and computational models of varying
complexity have been used to shed light on the macroscopic response of fibre-reinforced hydrogels.  
The Halpin--Tsai (HT) equations~\cite{halpin1976} are a 
relatively simple micromechanical model that 
describes
the stiffness of a fibre-reinforced matrix in terms of the
fibre and matrix properties.  Jordan \etal\cite{jordan2017} found
that the HT equations do not always provide accurate predictions
of the stiffness of fibre-reinforced hydrogels, and they attributed discrepancies to the large
contrast in the Young's moduli of the fibres and the hydrogel matrix. Beckett \etal\cite{beckett2024} also showed that the HT equations 
have limited applicability to fibre-reinforced hydrogels and proposed that swelling of the
hydrogel matrix could invalidate the assumptions that underpin
the HT equations.  
Visser \etal\cite{visser2015} developed
a simple expression for the stiffness of transversely isotropic fibre-reinforced hydrogels in terms
of the fibre properties.  Subsequent studies 
by Castilho \etal\cite{castilho2019} and Chen \etal\cite{chen2020}
used 
three-dimensional (3D) finite element simulations to understand, 
respectively, 
the nonlinear elastic and linear poroelastic response of the 
composites produced by Visser \emph{et al}.
Due to the specific geometry of the 
fibre network used by Visser \emph{et al}, in which the fibres
are laid down to form a two-dimensional (2D) grid, the 
modelling results of Castilho, Chen, \emph{et al.}\ might not
carry over to other systems.  Motivated by problems in cartilage
mechanics, Ateshian \etal\cite{ateshian2006, ateshian2009} developed a
general approach for modelling fibre-reinforced soft hydrated materials. However, simulating these models required
the development of a bespoke 3D nonlinear finite element code.  
The high computational cost of 3D finite element simulations, along
with the effort required to create bespoke code, reduces their use as predictive design tools.

\begin{figure}
\centering
\includegraphics[width=0.8\textwidth]{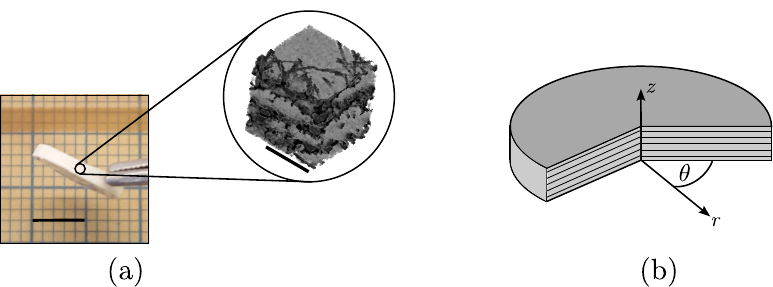}
\caption{(a) Photograph (scale bar = 5 mm) of a transversely isotropic fibre-reinforced hydrogel fabricated by Moore \etal\cite{moore2023} along with a 3D reconstruction using nano computed tomography (scale bar = 50 $\mu$m).
(b) A mathematical idealisation of a transversely isotropic fibre-reinforced hydrogel, with the layers of fibre network shown as parallel black
lines in the cross section.}
\label{fig:schematic}
\end{figure}

The purpose of this paper is, therefore,
to present a flexible and computationally efficient
microstructural modelling
framework for fibre-reinforced hydrogels that is implemented in a 
high-level open-source Python package. The modelling
framework links the
fibre properties and fibre network geometry to the macroscopic
material response.  We focus on modelling a class of
transversely isotropic materials developed by Moore \etal\cite{moore2023} to mimic the structure and mechanics of articular cartilage.  These materials
consisted of horizontally aligned fibre networks layered vertically and embedded in a hydrogel matrix, as shown in Fig.~\ref{fig:schematic}. 
Similar composites
have been fabricated by a number of other authors~\cite{strange2014, visser2015, bas2015, beckett2024}. 
We develop a microstructural model of the hydrogel and fibre phases, explicitly modelling the hydration of the gel, before considering unconfined compression tests of the composite material. We build upon techniques used for modelling fibre networks, and incorporate them within a nonlinear poroelastic framework that accounts for
osmotic effects.  The poroelastic modelling of the hydrogel
component of the material
follows standard approaches~\cite{hong2008, chester2010}; hence, the main focus of this paper is on modelling the fibre network.

Many studies have been conducted over the last several decades, producing models that have discovered key structural properties of fibre networks. 
Fibre networks are 3D in general, but can be modelled as 2D depending on their properties.  Niskanen and Alava \cite{niskanen1994planar} discovered that, at low coverage (the number of fibres covering a point in the plane), the transition from  2D to asymptotically 3D behaviour depends on the product of mean coverage and length; however, at high coverage, the transition depends on the product of the fibres' flexibility and width to thickness ratio. In terms of 2D network structure, the number of fibres per unit area defines a transition: above a critical density, percolation occurs (a contiguous set of fibres spans the entire network) \cite{li2009finite,li2013percolation}. Percolation is required for the network to behave as a solid. 

Fibre network \textit{structures} have been well characterised, but the focus in this paper is their mechanics. Early approaches to modelling network mechanics were based on regular or perturbed lattices \cite{schwartz1985behavior}, before being superseded by models in which straight fibres are laid randomly and isotropically in a 2D plane. This approach, which was developed for modelling paper and is now often called the ``Mikado'' model, has been used since at least the early 1990s. \r{A}str\"{o}m \etal \cite{aastrom1994microscopic} used it to predict the microscopic elastic and failure properties of high density networks and found that the effective Young's modulus of a network depends on both fibre density and length and that the stresses in the network follow an exponential distribution. This work built upon that of Cox \cite{cox1952elasticity}, who calculated how the stress in a matrix-embedded fibre varies along its length, and found that it can be described by a hyperbolic cosine function. However, it has been shown that the shear-lag type model used in Cox's paper does not apply to random fibre networks that are close to the percolation threshold \cite{raisanen1997does}. 

More recently, it has been shown that a fibre network's elasticity is affected not only by its connectivity, but also by the strain magnitude (the ``floppy-to-rigid transition") \cite{arzash2020finite,head2003distinct,shivers2019scaling}, and that its deformation falls into two regimes: one dominated by fibre bending and one by fibre stretching. In the stretching regime, the network transforms in an approximately affine way, whereas in the bending regime, the deformation is distinctly non-affine \cite{humphries2018mechanics}. The straightening and activation of fibre segments is critical to the mechanical behaviour of the network \cite{vainio2005observations,vainio2007interfiber}, a principle similar to the recruitment of collagen fibres in biological soft tissues \cite{shearer2015new,shearer2020recruitment,gregory2021microstructural,haughton2022bayesian}. In this paper, we consider two cases: firstly we assume that the fibre segments are straight and that the network deforms affinely, then we consider the gradual recruitment of wavy fibre segments. Rather than model the fibres as discrete objects, we use a continuum approach, which allows us, conveniently, to link our microscale fibre network models to a nonlinear, macroscale framework.

The paper is organised as follows.  In Sec.~\ref{sec:model},
we present the microstructural model for a transversely isotropic fibre-reincorced
hydrogel.  The parameter values
used in the model, which are based on the materials produced
by Moore \etal\cite{moore2023}, are described in Sec.~\ref{sec:params}.  In Sec.~\ref{sec:hydration} and Sec.~\ref{sec:unconfined}, the model is used to simulate
swelling and unconfined compression, respectively, and the model
predictions are compared to experimental data.  
In Sec.~\ref{sec:design}, we showcase how the model can be
used to guide the design of new materials that mimic the mechanical behaviour of artificial
cartilage.  Finally, the paper concludes in Sec.~\ref{sec:conclusion}.

\section{Model derivation}
\label{sec:model}

The governing equations are derived using the framework of
non-equilibrium thermodynamics. We begin by constructing
the free energy of the system, which accounts for the energy of
elastic deformations and hydrating the material.  The free energy
is then used to derive thermodynamically consistent constitutive
relations that can be used in the bulk equations.

\subsection{Microstructural modelling of the fibrous phase of the material}\label{micro}

We first derive two microstructural constitutive models for the fibrous phase of the material. We consider a two-dimensional network of fibres and assume that it deforms according to an affine mapping (an assumption that is appropriate for high-density fibre networks \cite{agarwal2023tensile}). Based upon this assumption, we calculate the strain energy in an arbitrary fibre segment with a specified initial angle, and then calculate the total strain energy in the network by integrating this expression over all angles. For the first constitutive model, we assume the fibres are all initially straight; in the second, we assume that they are wavy and are gradually recruited with increasing tensile strain.

\subsubsection{The strain energy function for fibre networks with initially straight fibres}

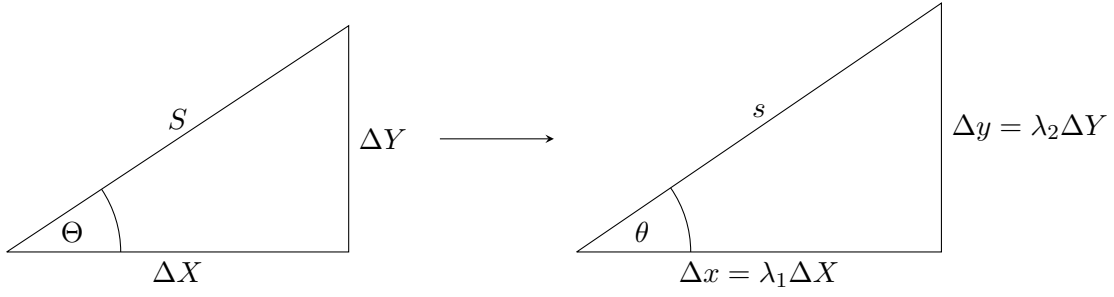
\begin{figure}[ht]
\centering
\begin{tikzpicture}[
  my angle/.style={
    every pic quotes/.append style=,
    draw,
    angle radius=1.5cm,
  },scale=1.5]
\coordinate [] (A) at (-1.5cm,-1.cm);
\coordinate [] (C) at (1.5cm,-1.0cm);
\coordinate [] (B) at (1.5cm,1.0cm);

\draw 
 (A) --
node[above] {$S$} (B) --
node[right] {$\Delta Y$} (C) -- 
node[below] {$\Delta X$} (A);

\pic [my angle, "$\Theta$"] {angle=C--A--B};

\draw [-stealth](2.3cm,0cm) -- (3.3cm,0cm);

\coordinate [] (a) at (3.5cm,-1.0cm);
\coordinate [] (c) at (6.7cm,-1.0cm);
\coordinate [] (b) at (6.7cm,1.2cm);

\draw 
 (a) --
node[above] {$s$} (b) --
node[right] {$\Delta y=\lambda_2\Delta Y$} (c) -- 
node[below] {$\Delta x=\lambda_1\Delta X$} (a);

\pic [my angle, "$\theta$"] {angle=c--a--b};
\end{tikzpicture}
\caption{The deformation of a fibre segment of initial length $S$ and initial angle $\Theta$.} \label{fig:fibre_segment}
\end{figure}

Consider a fibre segment of initial length $S$, that makes an initial angle $\Theta\in[0,\pi)$ with the $X$-axis, and is embedded in a fibre network of sufficiently high density that the network deforms in an affine manner. The fibre segment forms the hypotenuse of a right-angle triangle with initial side lengths $\Delta X$ and $\Delta Y$ aligned with the $X$- and $Y$-axes, respectively. Now assume that the network is subjected to an affine deformation consisting of homogeneous stretches $\lambda_1$ and $\lambda_2$ in the $X$- and $Y$-directions, respectively (without loss of generality, we shall assume $\lambda_1\ge\lambda_2$), so that the total deformation gradient is
\begin{align}
\tens{F}_t=\left(\begin{array}{cc}
\lambda_1 & 0\\
0 & \lambda_2
\end{array}\right).
\end{align}
We define the two-dimensional strain invariants associated with this deformation as $I_1=\lambda_1^2+\lambda_2^2$ and $I_2=\lambda_1^2\lambda_2^2$. The fibre segment is thus deformed according to Figure \ref{fig:fibre_segment}.
The following identities clearly hold:
\begin{align}
\Delta X=S\cos\Theta,\quad\Delta Y=S\sin\Theta,
\end{align}
\begin{align}  
s^2=\Delta x^2+\Delta y^2=\lambda_1^2\Delta X^2+\lambda_2^2\Delta Y^2=S^2\lambda_1^2\cos^2\Theta+S^2\lambda_2^2\sin^2\Theta.
\end{align}
Thus, the stretch in the fibre segment, $\lambda$, is given by
\begin{align}
\lambda=\frac{s}{S}=\sqrt{\lambda_1^2\cos^2\Theta+\lambda_2^2\sin^2\Theta}.
\label{stretch}
\end{align}
We shall assume that the constitutive behaviour of each individual fibre segment is Hookean so that its strain energy density (per unit \textit{initial} volume) function is
\begin{align}
W_f^{(i)}(\lambda)=\frac{E_f}{2}(\lambda-1)^2,
\label{W}
\end{align}
where $E_f$ is the fibre Young's modulus. 
Upon substituting \eqref{stretch} into \eqref{W}, we can write the strain energy density in the $i^\text{th}$ fibre segment as a function of the initial fibre angle:
\begin{align}
W_{f}^{(i)}(\Theta)=\frac{E_f}{2}\left(\sqrt{\lambda_1^2\cos^2\Theta+\lambda_2^2\sin^2\Theta}-1\right)^2.
\end{align}


Note that we have considered the strain energy \textit{density} so far. The strain energy stored in the $i^\textrm{th}$ fibre segment will be $U_f^{(i)}=V_f^{(i)}W_f^{(i)}$, where $V_f^{(i)}$ is its volume. Therefore, the total energy stored in the fibre network is given by $U_f^\text{tot} =\sum_{i=1}^N U_f^{(i)}$, where $N$ is the number of fibre segments in the network. This can be written as $U_f^\text{tot}=N \bar{U}_f$, where $\bar{U}_f =\mathbb{E}[U_f^{(i)}]$ is the average amount of strain energy stored in each fibre segment, i.e.\ the \textit{expectation} of $U_f^{(i)}$. Since $V_f^{(i)}$ and $W_f^{(i)}$ are \textit{independent} random variables for the affine deformation we are considering, we can write $\bar{U}_f = \bar{V}_f \bar{W}_f$, where $\bar{V}_f = \mathbb{E}[V_f^{(i)}]$ and $\bar{W}_f = \mathbb{E}[W_f^{(i)}]$. The total volume occupied by fibres is $V_f^\text{tot} =N \bar{V}_f$; therefore, the total energy stored in the network is $U_f^\text{tot}=V_f^\text{tot} \bar{W}_f$. To obtain the total strain energy \textit{density} of the whole fibre network, we simply divide through by the volume of the solid in which the fibres are embedded, $V_s$, to give $(V_f^\mathrm{tot}/V_s)\bar{W}_f=\Phi_f \bar{W}_f$, where $\Phi_f$ is the volume fraction of the fibre segments that are mechanically active in the network.  
In the following, we shall drop the bar on $\bar{W}_f$ for
convenience.

We assume that there are so many fibres in the network that we can represent the distribution of their initial angles using a continuous probability density function. Since the fibres are isotropically oriented in the plane
we shall assume a uniform probability density $\mathcal{U}_{[0,\pi]}(\Theta)$. Therefore, the average strain energy density is
\begin{align}
W_f=\int_{-\infty}^\infty \mathcal{U}_{[0,\pi]}(\Theta)W_f^{(i)}(\Theta)\,\d\Theta=\frac{1}{\pi}\int_0^\pi W_f^{(i)}(\Theta)\,\d\Theta.
\label{WTheta}
\end{align}
Upon evaluating this integral, we obtain
\begin{align}
W_f&=\frac{E_f}{4}\left(\lambda_1^2+\lambda_2^2-\frac{8\lambda_1}{\pi}\mathcal{E}\left(1-\frac{\lambda_2^2}{\lambda_1^2}\right)+2\right)\label{Wflambda}\\
&=\frac{E_f}{4}\left(I_1-\frac{4\sqrt{2}}{\pi}\sqrt{I_1+\sqrt{I_1^2-4I_2}}\mathcal{E}\left(2\frac{\sqrt{I_1^2-4I_2}}{\sqrt{I_1^2-4I_2}+I_1}\right)+2\right),
\label{Wf}
\end{align}
where $\mathcal{E}(\cdot)$ is the complete elliptic integral of the second kind. Equation \eqref{Wflambda} is only valid for $\lambda_1\ge\lambda_2$; when $\lambda_2>\lambda_1$, the positions of $\lambda_1$ and $\lambda_2$ in the function are swapped. In either case, the expression in terms of the invariants \eqref{Wf} holds.

To express this two-dimensional strain energy function in three dimensions, we assume that the layered structure of the material gives rise to transversely isotropic material behaviour, with the $Z$-axis being the axis of anistropy, and we assume that $\lambda_3$ is the principal stretch in that direction. For such a material, there are five strain invariants. The following set of invariants is commonly used:
\begin{align}
\hat{I}_1=\text{tr}(\hat{\tens{C}})=\lambda_1^2+\lambda_2^2+\lambda_3^2,\quad\hat{I}_2=\frac{1}{2}(I_1^2-\text{tr}(\hat{\tens{C}}^2))=\lambda_1^2\lambda_2^2+\lambda_1^2\lambda_3^2+\lambda_2^2\lambda_3^2,
\end{align}
\begin{align}
\hat{I}_3=\det(\hat{\tens{C}})=\lambda_1^2\lambda_2^2\lambda_3^2,\quad\hat{I}_4=\vec{M}\cdot(\hat{\tens{C}}\vec{M})=\lambda_3^2,\quad\hat{I}_5=\vec{M}\cdot(\hat{\tens{C}}^2\vec{M})=\lambda_3^4,
\end{align}
where $\hat{\tens{C}}=\hat{\tens{F}}^T\hat{\tens{F}}$, is the three-dimensional right Cauchy--Green deformation tensor, $\hat{\tens{F}}$ is the three-dimensional deformation gradient, $\vec{M}=\vec{E}_Z$ is a unit vector pointing in the direction of the axis of anisotropy in the initial configuration, and the latter two equations make use of the assumption that the deformation gradient has no shear strains involving $Z$ so that $\hat{F}_{13}=\hat{F}_{23}=\hat{F}_{31}=\hat{F}_{32}=0$. Therefore, by comparing with the definitions of the 2D invariants, $I_1$ and $I_2$, we can see that
\begin{align}
    I_1=\hat{I}_1-\hat{I}_4,\quad I_2=\frac{\hat{I}_3}{\hat{I}_4},
    \label{eqn:2D_3D}
\end{align}
which can be substituted into equation \eqref{Wf} to obtain
\begin{multline}
 W_f=\frac{E_f}{4}\Biggl(
\hat{I}_1-\hat{I}_4\\
\left.
-\frac{4\sqrt{2}}{\pi}\sqrt{\hat{I}_1-\hat{I}_4+\sqrt{(\hat{I}_1-\hat{I}_4)^2-4\hat{I}_3/\hat{I}_4}}\mathcal{E}\left(2\frac{\sqrt{(\hat{I}_1-\hat{I}_4)^2-4\hat{I}_3/\hat{I}_4}}{\sqrt{(\hat{I}_1-\hat{I}_4)^2-4\hat{I}_3/\hat{I}_4}+\hat{I}_1-\hat{I}_4}\right)+2\right).
\end{multline}

\subsubsection{The strain energy function for fibre networks with initially wavy fibres}

The fibre strain energy density function above was for initially straight fibre segments. Now, we will assume that the fibre segments are wavy and only resist tension once the excess length has been straightened out. We assume that there is a \textit{distribution} of waviness and we are interested in calculating the \textit{average} strain energy density of a fibre segment for a given waviness distribution. We characterise the waviness of an individual fibre segment in terms of a critical recruitment stretch, $\lambda_c$, and assume that its strain energy density function is
\begin{align}
W_f^{(i)}(\lambda,\lambda_c)=\frac{E_f}{2}\left(\frac{\lambda}{\lambda_c}-1\right)^2\mathcal{H}(\lambda-\lambda_c),
\end{align}
where $\mathcal{H}\mathcal(\cdot)$ is the Heaviside function. This represents a fibre that does not store any energy if the current stretch is below its recruitment stretch and is linear elastic once the stretch is above its recruitment stretch. The \textit{average} strain energy density (averaged over the critical recruitment stretch distribution function $f(\lambda_c)$) is
\begin{align}
    \tilde{W}_f(\lambda)=\int_0^\infty f(\lambda_c)W_f^{(i)}(\lambda,\lambda_c)\,\d\lambda_c=\frac{E_f}{2}\int_0^\lambda f(\lambda_c)\left(\frac{\lambda}{\lambda_c}-1\right)^2\,\d\lambda_c.
    \label{W(lambda)}
\end{align}
Following the arguments above, this can then be expressed as a function of initial angle as
\begin{align}
    \tilde{W}_f(\Theta)=\frac{E_f}{2}\bigintss_0^{\sqrt{\lambda_1^2\cos^2\Theta+\lambda_2^2\sin^2\Theta}} f(\lambda_c)\left(\frac{\sqrt{\lambda_1^2\cos^2\Theta+\lambda_2^2\sin^2\Theta}}{\lambda_c}-1\right)^2\,\d\lambda_c.
\end{align}
Then, the strain energy density in the whole network (again assuming a uniform distribution of initial fibre orientations) can be written as:
\begin{align}
W_f&=
\bigintss_{-\infty}^\infty \mathcal{U}_{[0,\pi]}(\Theta)\tilde{W}_f(\Theta)\,\d\Theta\nonumber\\
&=\frac{E_f}{2\pi}\bigintss_0^\pi\bigintss_0^{\sqrt{\lambda_1^2\cos^2\Theta+\lambda_2^2\sin^2\Theta}}f(\lambda_c)\left(\frac{\sqrt{\lambda_1^2\cos^2\Theta+\lambda_2^2\sin^2\Theta}}{\lambda_c}-1\right)^2\,\d\lambda_c\,\d\Theta.
\label{Wtot}
\end{align}
These integrals can be evaluated 
for a given recruitment stretch distribution, $f(\lambda_c)$, and given values of the principal stretches, $\lambda_1$ and $\lambda_2$.  In this work, we assume
the recruitment stretch distribution is
given by a piecewise quartic polynomial
\begin{align}
    f(\lambda_c)=\begin{cases}
        0, & \lambda_c\le1,\\
        \dfrac{60\lambda_c^2(\lambda_c-1)(\lambda_c-\lambda_m)}{3-5\lambda_m+5\lambda_m^4-3\lambda_m^5}, & 1\le\lambda_c\le\lambda_m,\\
        0, & \lambda_c\ge\lambda_m,
    \end{cases}
\label{eqn:quartic}
\end{align}
where $\lambda_m$ is the maximum recruitment
stretch required to engage every fibre in the
network.  The quartic distribution  allows the double
integral in \eqref{Wtot} to be calculated exactly, albeit in terms of
a lengthly expression involving
special functions (see Appendix \ref{app} for details).

We find that it is sometimes more computationally efficient to use
a semi-analytical approach where
the inner integral in \eqref{Wtot} is computed exactly
and the outer integral is numerically
approximated using the 
trapezoidal rule, which is exponentially accurate due to the periodicity of the integrand~\cite{trefethen2014}.  This hybrid approach allows 
a variety
of fibre recruitment stretch distributions $f(\lambda_c)$
to be used in the model (e.g.\ uniform, normal, beta, gamma, exponential)~when an exact expression for the 
double integral in
\eqref{Wtot} is not available. However, in the implementation discussed below, we use the piecewise quartic function defined in equation \eqref{eqn:quartic}.


\subsection{Combining the fibrous phase with the isotropic hydrogel matrix}\label{combining}

In the previous section, we derived the strain energy function for the fibrous phase of the material. For the hydrogel matrix phase, we assume that the constitutive behaviour is described by an isotropic, compressible, neo-Hookean strain energy function:
\begin{align}
  W_m = \frac{E_m}{4(1+\nu_m)}\left(\hat{I}_1 - 3 - \log \hat{I}_3\right)+\frac{E_m\nu_m}{(1+\nu_m)(1-2\nu_m)}\left(\sqrt{\hat{I}_3}-1\right)^2,
\end{align}
where 
$E_m$ and $\nu_m$ are the Young's modulus and Poisson's ratio of the matrix, respectively. The total hyperelastic strain energy function of the elastic phase is then
\begin{align}
  W_e(\hat{I}_1,\hat{I_3},\hat{I}_4) = (1 - \Phi_f) W_m(\hat{I}_1,\hat{I_3}) + \Phi_f W_f(\hat{I}_1,\hat{I_3},\hat{I}_4),
  \label{eqn:W_s}
\end{align}
where we recall that $\Phi_f$ represents the nominal\footnote{We use the word nominal to describe quantities in the non-hydrated configuration of the material.} volume fraction of fibres in the non-hydrated state. From this point onward, we drop the hat notation for three-dimensional quantities, since we will no longer need to refer to the two-dimensional theory.

Now that we have derived the strain energy function for the solid phase of the material, we move on to modelling energy changes associated with its hydration.

\subsection{Material hydration}

To model the energy change of hydration, we assume the material can be treated as a single homogenised body characterised by the nominal fibre fraction $\Phi_f$
and the nominal water fraction $\Phi_w$.  The nominal water fraction measures the volume of water per unit of undeformed (non-hydrated) volume.  The Flory--Huggins theory of 
solvent-polymer mixtures~\cite{doi1996} is used to describe the energy of
mixing water molecules, polymer chains, and fibre segments.  The
free energy density (per unit undeformed volume) is given by
\begin{align}
W_\text{mix} = \frac{R_g T}{\Omega_w}\left[\Phi_w \log \left(\frac{\Phi_w}{1 +\Phi_w}\right) + \frac{\chi(\Phi_f) \Phi_w}{1 + \Phi_w}\right],
\label{eqn:W_mix}
\end{align}
where $\Omega_w = 18 \times 10^{-6}$ m$^3$ mol$^{-1}$ is the molar volume of water, $R_g = 8.314$ J mol$^{-1}$ K$^{-1}$ is the universal gas constant, $T = 296$ K is temperature, and $\chi$ is the Flory interaction parameter, which characterises the strength of unfavourable interactions between water molecules and solid components (polymers and fibres).  If $\chi \gg 1$, then it is energetically costly to hydrate the material; in this case, the amount of water uptake during the hydration phase will be small. 
The dependence of $\chi$ on the nominal fibre fraction $\Phi_f$
reflects the differing affinity of water to the fibre and matrix phases. 
Its functional form will be discussed in Section \ref{sec:case_study}.

\subsection{Balance laws and bulk equations}
\label{sec:balance_laws}

\begin{figure}
\centering
\includegraphics[width=0.85\textwidth]{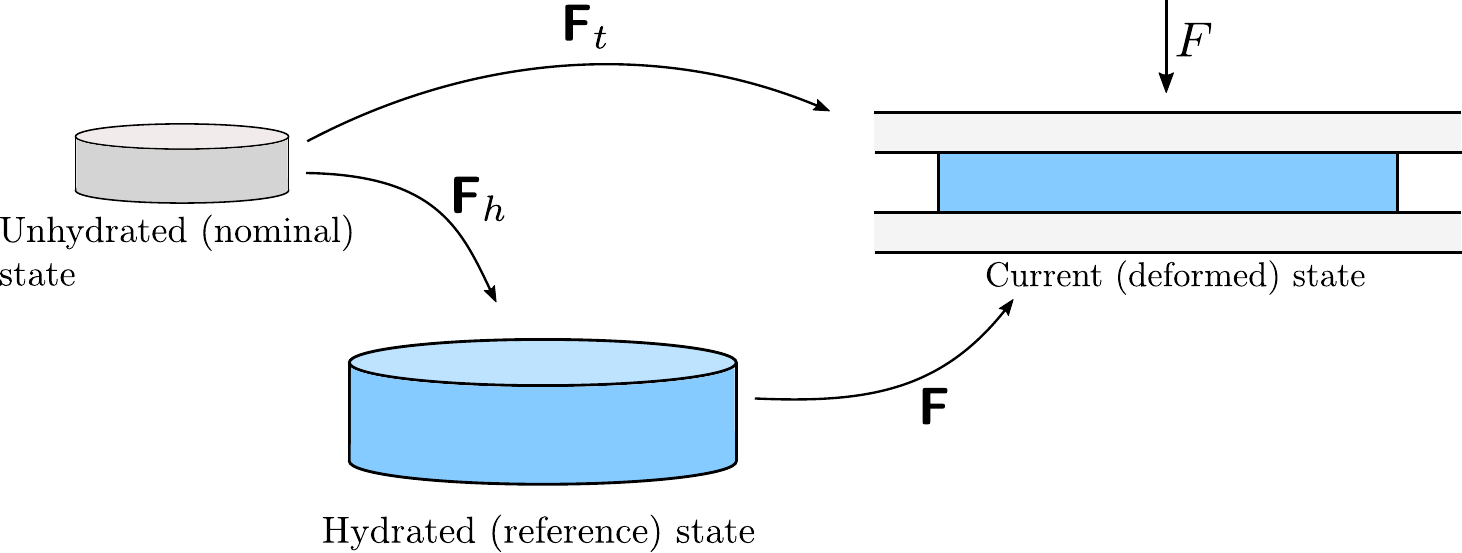}
\caption{The three spatial configurations used
in the model and the deformation gradient
tensors used to move between them.  The unhydrated (nominal) state represents the material as it was prepared before hydration. The
hydrated (reference) state corresponds to a freely swollen material and is used as a reference
configuration to formulate the model equations.
The current (deformed) state in this work 
corresponds to an unconfined
compression experiment, where a cylindrical 
sample has been placed between two plattens
and an axial force, $F$, has been applied.}
\label{fig:configurations}
\end{figure}

Having defined the free energy density of the material, $W = W_e + W_\text{mix}$, the 
constitutive relations for a three-dimensional,
time-dependent model can now be derived using
non-equilibrium thermodynamics. For brevity, we only
present the most important equations here; their
derivations are provided in Appendix~\ref{sec:thermo}. 
We choose to formulate the governing equations using
a reference configuration 
associated with a freely swollen,
hydrated state (to be determined in Section~\ref{sec:hydration}). Here, freely swollen means that there are no geometric constraints or loads applied to the material.   
Thus, we let the deformation gradient, $\tens{F}_h$, describe how the non-hydrated material
deforms upon hydration; see
Fig.~\ref{fig:configurations}. 
The relative volume change due to hydration is given by $J_h = \det \tens{F}_h$. We assume that the volume of the hydrated material, $V_h$, is equal to the sum of the initial volume of the solid phase, $V_s$, and the added volume of water, $V_w$. Therefore, $V_h=V_s+V_w$ and $J_h=V_h/V_s=1/(1-\phi_0)$, where $\phi_0=V_w/V_h$ is the initial volume fraction of water in the hydrated state.  That is, $\phi_0$ represents
the initial porosity of the material. 

Subsequent deformations relative to the hydrated state are described by the deformation gradient tensor $\tens{F}$.  By introducing the displacement
relative to the hydrated state, $\vec{u}$, the deformation
gradient tensor can be written as
$\tens{F} = \tens{I} + \nabla \vec{u}$, where
$\tens{I}$ is the identity tensor and $\nabla$ is
the gradient with respect to coordinates in the
hydrated state.
Similarly, $J = \det \tens{F}$ describes the volume of a material
element in the current (deformed) state relative to its volume
in the hydrated (reference) state. 
The total deformation gradient tensor that describes
the deformation from the non-hydrated (nominal) 
configuration to the current (deformed)
configuration is given by $\tens{F}_t = \tens{F} \tens{F}_h$. The
total volume change of a material element 
relative to its volume in the non-hydrated state is
$J_t = \det \tens{F}_t = J J_h$.

The constituents of the material (water, polymers, fibres) are 
assumed to be incompressible.  However, at the macroscopic level,
the material can compress or expand due to local rearrangements of the polymer and fibre networks.  These rearrangements are driven by
the imbibition of water into, or the expulsion of water out of,
material elements.  Thus, the volume
of a given material element is dependent on the volume of water
contained within it.  Since $J_t$ describes the volume of material (water and solid components) per unit unhydrated volume,
and $\Phi_w$ is the volume of water per unit unhydrated volume, 
we can write
$J_t = 1 + \Phi_w$, a condition known as the molecular
incompressibility condition. By introducing the
referential water fraction $\phi_w = \Phi_w / J_h$, 
which measures the volume of
water per unit reference volume (the volume of the initial hydrated state), 
the incompressibility condition can be expressed as
\begin{align}
J = 1 + \phi_w - \phi_0.
\end{align}
The current volume fraction of water (volume of water per unit current volume, also known as the porosity) is
defined as $\phi = \phi_w / J$.

Conservation of mass in the water phase implies that the referential fraction of water, $\phi_w$, evolves according to
\begin{align}
  \pd{\phi_w}{t} + \nabla \cdot \vec{Q}_w = 0,
\end{align}
where $t$ is time and $\vec{Q}_w$ is the referential volumetric flux of water (volume of water per unit time per unit area of the hydrated state).
The flux of water is given by a Lagrangian form of Darcy's law,
\begin{align}
  \vec{Q}_w = -k(J) J \tens{C}^{-1} \nabla p,
  \label{eqn:Q_w}
\end{align}
where $k(J)$ is the deformation-dependent isotropic 
permeability of the material (divided by the
water viscosity) in the current state, $\tens{C} = \tens{F}^T \tens{F}$ is the right Cauchy--Green tensor, and $p$ is the
water pressure.  The permeability in the current configuration is likely not isotropic in reality due to the anisotropy of the material being modelled, but the assumption of isotropy is sufficient for our purposes as we will later simplify the model by postulating that water transport only occurs in the radial direction.
We use the Holmes--Mow permeability equation~\cite{holmes1990} given by
\begin{align}
k(J) = k_0 \left(\frac{J - \phi_0}{1 - \phi_0}\right)^\gamma \exp\left[\frac{M}{2}(J^2 - 1)\right],
\label{eqn:perm}
\end{align}
where $k_0$ is the permeability of the hydrated material
and $\alpha$ and $M$ are fitting parameters. 



Conservation of linear momentum, neglecting inertial terms, leads to
\begin{align}
  \nabla \cdot \tens{S} = \mathbf{0},
\end{align}
where $\tens{S}$ is the referential stress tensor, which
measures forces per unit initial area of the hydrated state. The total stress tensor is the sum of the elastic
stress $\tens{S}^e$, the osmotic stress $\Pi$, and the
fluid pressure $p$, such that 
\begin{align}
\tens{S} = \tens{S}^e - [\Pi(\phi_w) + p] J \tens{F}^{-T}.  
\label{eqn:S}
\end{align}
The elastic component of the referential stress tensor is given by
\begin{align}
\tens{S}^e &= \frac{1}{J_h}\pd{W_e}{\tens{F}_t}\tens{F}_h^{T},
\end{align}
where the strain energy density $W_e$ is given by \eqref{eqn:W_s}
and $\left(\dfrac{\partial}{\partial\tens{A}}\right)_{ij}=\dfrac{\partial}{\partial A_{ij}}$ for any second-order tensor $\tens{A}$. 
In the absence of any deformation from the initial, hydrated state, so that
$\tens{F} = \tens{I}$ and $J = 1$, the tensor $\tens{S}$ gives the \textit{Cauchy} stress in the initial, hydrated state.
The osmotic stress is derived from the Flory--Huggins mixing
energy~\eqref{eqn:W_mix} and given by
\begin{align}
\Pi(\phi_w) = -\pd{W_\text{mix}}{\Phi_f} = -\frac{R_g T}{\Omega_w} \left[\log \left(\frac{\phi_w}{J}\right) + 1 - \frac{\phi_w}{J} + \chi(\Phi_f) \left(1 - \frac{\phi_w}{J}\right)^2\right].
\end{align}

The initial condition for this system of equations is 
that the referential water fraction $\phi_w$ is equal to
the porosity of the initially hydrated state,
$\phi_0$. The boundary conditions to be imposed are specific
to the problem being considered and will be discussed in the
context of free swelling and unconfined compression experiments
in Sec.~\ref{sec:hydration} and Sec.~\ref{sec:unconfined} below.

\section{Material specification and parameter values}
\label{sec:params}

We consider materials that are based on the
fibre-reinforced hydrogels fabricated by Moore \etal\cite{moore2023} by electrohydrodynamic deposition
(electrospinning and electrospraying).
The authors referred to their
materials as `FiHy'.  The hydrogel matrix of FiHy
is composed
of gelatin ($E_m = 4$~kPa, $\nu_m = 0$, $\rho_m = 1.2$ g cm$^{-3}$). 
The fibres are composed of 
PCL ($E_f = 400$~MPa, 
$\rho_f = 1.1$ g cm$^{-3}$).  
These parameters values are obtained from
Refs.~\cite{moore2023, chembook, beckett2024}.
The nominal fibre fractions ranged from 
0.25 to 0.76 \emph{by mass}.  The nominal
fibre fraction by volume is calculated to range from
$\Phi_f = 0.26$ to $0.77$. The FiHy samples were hydrated for
24 hours with 1X phosphate buffered saline (PBS), 
which has a density that is approximately equal to
that of water ($\rho_w = 1$ g cm$^{-3}$).  After hydration, the samples were approximately 80\% water by mass, regardless of the fibre fraction.  The volume fraction of water in the hydrated state can be calculated as $\phi_0 \simeq 0.82$ for all
fibre fractions.

\section{Hydration of a fibre-reinforced hydrogel}
\label{sec:hydration}

Now that we have derived our governing equations, we explicitly calculate the freely swollen states that are achieved upon hydration of the non-hydrated material.  For this, we envision that
the material has been submersed in
a bath of water.  Equilibrium will
be reached when the total stress $\tens{S}$
and the fluid stress (pressure) $p$ within
the composite balance the
pressure of the surrounding fluid,
which can be set to zero without
loss of generality.  Therefore, upon
full hydration,
the total stress and the fluid pressure within the gel 
will be zero, $\tens{S} = \tens{0}$ and $p = 0$.
Using the expression for $\tens{S}$ given by
\eqref{eqn:S}, these equations can be combined into
\begin{align}
\tens{S}^e = \Pi(\phi_0) \tens{I},
\label{eqn:S_0}
\end{align}
where we have used $\tens{F} = \tens{I}$ for the hydrated state
and we recall from Sec.~\ref{sec:balance_laws} that $\phi_0 = 1 - 1 / J_h$ is the volume fraction of fluid in the initial
hydrated state. Equation \eqref{eqn:S_0} implies that in a free-swelling scenario, the elastic and osmotic stresses must
balance.  Moreover, since $\Pi(\phi_0) > 0$, 
it follows that the osmotic stress
is tensile and acts to stretch the material.

We assume that the hydration of the sample leads to anisotropic, but homogeneous swelling of the solid phase with corresponding deformation gradient
\begin{align}
\tens{F}_h=\left(\begin{array}{ccc}
    \alpha_\parallel & 0 & 0 \\
    0 & \alpha_\parallel & 0 \\
    0 & 0 & \alpha_\perp
    \end{array}\right)\quad\Rightarrow\quad J_h=\alpha_\parallel^2\alpha_\perp,
    \label{eqn:Fh}
\end{align}
where $\alpha_\parallel$ and $\alpha_\perp$ are stretches in the direction
parallel and perpendicular to the plane of the fibres, respectively.
The form of \eqref{eqn:Fh} leads to an elastic stress tensor of the form
\begin{align}
    \tens{S}^e=\left(\begin{array}{ccc}
    \sigma^e_\parallel(\alpha_\parallel,\alpha_\perp) & 0 & 0 \\
    0 & \sigma^e_\parallel(\alpha_\parallel,\alpha_\perp) & 0 \\
    0 & 0 & \sigma^e_\perp(\alpha_\parallel,\alpha_\perp)
    \end{array}\right),
    \label{eqn:Se}
\end{align}
with $\sigma^e_\parallel$ and $\sigma^e_\perp$ denoting the in-plane and out-of-plane Cauchy stresses of the hydrated state, which are calculated in Appendix \ref{app:equibiaxial} and are given by \eqref{eqn:sigma_para}--\eqref{eqn:sigma_perp}.
For a given value of the Flory interaction parameter, $\chi$, 
equations \eqref{eqn:S_0}--\eqref{eqn:Fh} and provide three independent equations for the three unknowns: $\alpha_\parallel$, $\alpha_\perp$,
and $J_h$.  Alternatively, given an experimentally observed value of $J_h$,
these three equations determine $\alpha_\parallel$, $\alpha_\perp$, and $\chi$. 

\subsection{Case study: swelling of PCL-reinforced gelatin}
\label{sec:case_study}

To illustrate the behaviour of the model, we numerically solve \eqref{eqn:S_0} and \eqref{eqn:Fh} 
to explore how the hydration of a fibre-reinforced hydrogel depends on
the Flory interaction parameter and the nominal fibre fraction.  Calculating the degree of hydration
requires values for the mechanical properties of
the fibre and matrix; these are taken from
Sec.~\ref{sec:params}.



\begin{figure}
\centering
\includegraphics[width=0.99\textwidth]{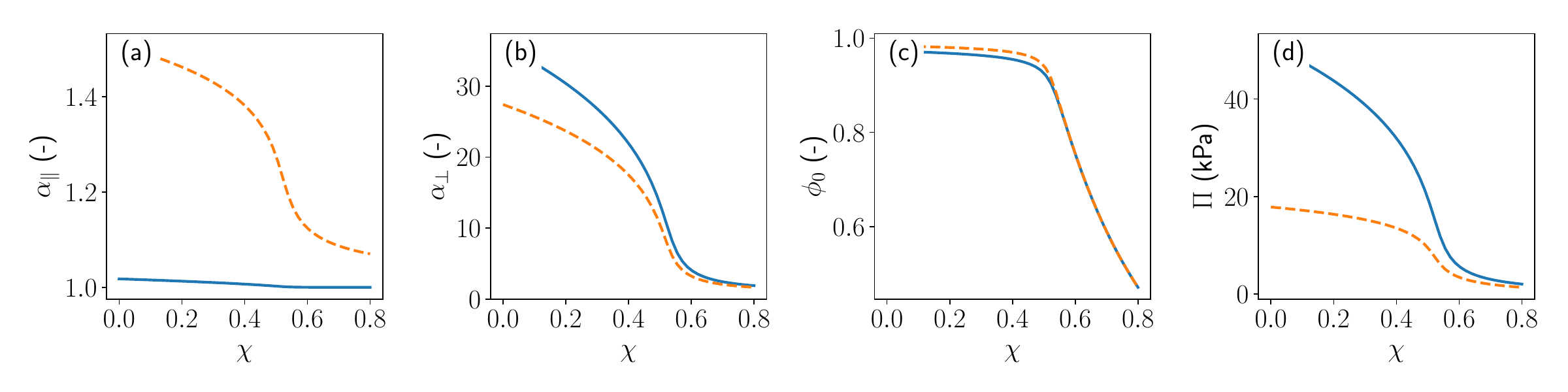}
\includegraphics[width=0.99\textwidth]{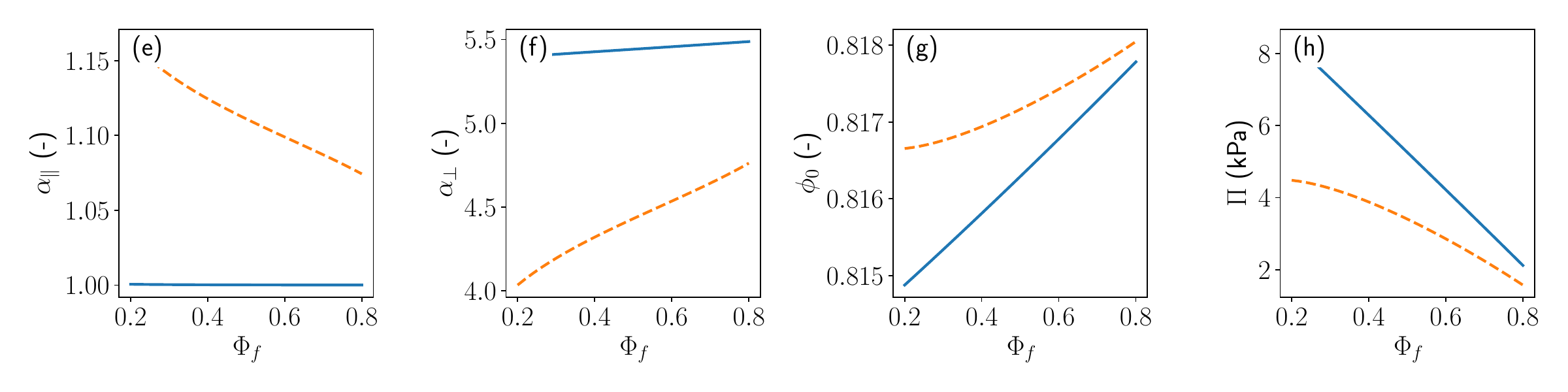}
\caption{Hydration of fibre-reinforced hydrogels without ($\lambda_m = 1$, solid lines) and with ($\lambda_m = 3$, dashed lines) fibre recruitment.  The panels show the in-plane stretch (a, e), out-of-plane stretch (b, f), porosity of the hydrated state (c, g), and the osmotic stress (d, h) as a function of the Flory interaction parameter, $\chi$, with $\Phi_f=0.26$ (a-d), and the nominal fibre fraction, $\Phi_f$, with $\chi = 0.57$ (e-h).}
\label{fig:hydration}
\end{figure}

We first fix the nominal volume fraction of fibres
to $\Phi_f = 0.26$ and vary the Flory interaction
parameter $\chi$.  We consider fibre networks
that are initially fully engaged with $\lambda_m = 1$
and partially engaged with $\lambda_m = 3$ (recall that $\lambda_m$ is the maximum recruitment stretch in the distribution function \eqref{eqn:quartic}).
In both cases, the deformation due to hydration
is strongly anisotropic, with the radial stretch
being much smaller than the axial stretch;
see Fig.~\ref{fig:hydration}~(a)--(b).  
The anisotropic swelling is
caused by the large contrast between the Young's moduli of the fibres and the matrix. 
As the Flory interaction parameter increases,
the radial and axial stretches decrease because the
volume of water absorbed by the matrix decreases,
which is reflected in the porosity of the hydrated
state strongly decreasing with $\chi$ (Fig.~\ref{fig:hydration}~(c)).  
Moreover, with increasing $\chi$, 
it becomes less energetically favourable to swell
the network and hence the osmotic stress decreases
(Fig.~\ref{fig:hydration} (d)). 
In the case of a fully engaged fibre network ($\lambda_m = 1$),  
there is virtually no in-plane expansion of the material and all of the deformation is 
perpendicular to the plane of the fibres.  
The small in-plane
deformation is expected because the osmotic stress responsible for hydration,
$\Pi \sim 10$~kPa, 
is much smaller than the in-plane Young's modulus, 
which is dominated by the
fibre modulus $E_f \sim 400$ MPa.  Conversely, the out-of-plane
Young's modulus is controlled by the gelatin matrix, $E_m \sim 4$~kPa, and is comparable to the osmotic stress; hence, there is a substantial 
out-of-plane deformation.  When the fibre network
is only partially engaged ($\lambda_m = 3$),
we see there is a larger in-plane
deformation, which is due to initially inactive fibres causing a reduction in the in-plane stiffness of the material, and a smaller out-of-plane deformation.  However, the in-plane stretching caused
by hydration is always less than the maximum recruitment stretch.  Therefore, hydration is not
sufficient to fully engage the fibre network. 
Remarkably, the out-of-plane deformation decreases in such a way that the porosity of the material remains approximately the same in both cases.  Thus, the
degree of hydration is not significantly impacted
by the initial level of fibre engagement.

To study how material hydration depends on
the nominal fibre fraction, a model for
the Flory interaction parameter $\chi$ 
in \eqref{eqn:W_mix} is required.  The
experiments of Moore~\etal\cite{moore2023} 
showed that after 24 hours of hydration, the materials had a 
porosity of $\phi_0 \simeq 0.82$ across a range of nominal
fibre volume fractions ($\Phi_f = 0.26$ to $0.77$).
These findings suggest that the presence of
fibres does not strongly impact the Flory
interaction parameter.  Thus, we set $\chi$ to
a constant given by $\chi = 0.57$, which
is determined by reading off the value of $\chi$
in Fig.~\ref{fig:hydration}~(c) that leads to 
a porosity of $\phi_0 = 0.82$.


By fixing $\chi = 0.57$ and considering
fibre networks that are initially fully
engaged ($\lambda_m = 1$), we see that the deformation and porosity do 
not strongly depend on the nominal
fibre fraction (Fig.~\ref{fig:hydration} (e)--(g)). The osmotic stress decreases
linearly with $\Phi_f$ (Fig.~\ref{fig:hydration} (h)), reflecting
the balance between osmotic stress
and the out-of-plane elastic stress produced
by the gelatin matrix, the latter of which
is proportional to $1 - \Phi_f$; see Appendix \ref{app:equibiaxial}.  When initially inactive fibres are accounted for ($\lambda_m = 3$), hydration
leads to greater in-plane stretching
and reduced out-of-plane stretching 
compared to the active case (as is the case when these parameters are plotted as a function of $\chi$).  Moreover, both the in-plane and out-of-plane stretches now strongly depend on the fibre fraction $\Phi_f$. The dependence
of the stretches on $\Phi_f$ ultimately results from the in-plane softening that
occurs due to inactive fibres. This
softening allows the osmotic stress to
drive a greater in-plane expansion of the material.
As the nominal fibre fraction increases, the in-plane stiffness increases as well,
reducing the in-plane deformation and increasing the out-of-plane deformation. Again, whether the fibres are initially active or inactive does not affect the porosity as a function of $\Phi_f$ (Fig. \ref{fig:hydration} (g)). 



\section{Modelling unconfined compression experiments}
\label{sec:unconfined}

We consider a cylindrical sample of hydrated material
of initial radius $R$, height $H$, and porosity
$\phi_0$. The sample is subjected to an unconfined compression along its longitudinal axis by placing it between two impermeable plates, which are then compressed via an imposed displacement or
force $F$, as shown in Fig.~\ref{fig:configurations}.  In this experiment, the sample is surrounded by water to prevent dry out.
During compression, the sample is free to expand in the radial direction.  The plates are assumed to be frictionless so that the cylindrical shape of the sample is preserved during
compression. Thus, the height of the sample decreases to $h(t)$
and its radius increases to $r_c(t)$. The deformation gradient then takes the form
\begin{align}
    \tens{F}=\left(\begin{array}{ccc}
    \beta_r(r,t) & 0 & 0 \\
    0 & \beta_\theta(r,t) & 0 \\
    0 & 0 & \beta_z(t)
    \end{array}\right)\quad\Rightarrow\quad J(r,t)=\beta_r(r,t)\beta_\theta(r,t)\beta_z(t),
    \label{eqn:F_cylinder}
\end{align}
where $\beta_r$, $\beta_\theta$, and $\beta_z$ are
the principal stretches in the radial, circumferential, and
axial direction, respectively. The coordinate
$r$ measures radial distances in the initially hydrated state; thus, $0 \leq r \leq R$.  
The total stretches, measured relative to the undeformed
(non-hydrated) state are given by
$\lambda_r = \alpha_\parallel \beta_r$,
$\lambda_\theta = \alpha_\parallel \beta_\theta$, and
$\lambda_z = \alpha_\perp \beta_z$.
The principal (referential) stresses are given by
$S_i = S_i^e - [\Pi(\phi_w) + p] \beta_i^{-1} J$ with 
\begin{align}
S_r^e = \frac{\alpha_\parallel}{J_h}\pd{W_e}{\lambda_r}, 
\quad
S_\theta^e = \frac{\alpha_\parallel}{J_h}\pd{W_e}{\lambda_\theta}, 
\quad
S_z^e = \frac{\alpha_\perp}{J_h}\pd{W_e}{\lambda_z}.
\label{eqn:S_i}
\end{align}
In \eqref{eqn:S_i}, the derivatives of $W_e$ are evaluated
at $\lambda_r = \alpha_\parallel \beta_r$,
$\lambda_\theta = \alpha_\parallel \beta_\theta$, and
$\lambda_z = \alpha_\perp \beta_z$.

\subsection{Reduction of the governing equations}

In light of the form of the deformation gradient
tensor in \eqref{eqn:F_cylinder}, we introduce the
radial displacement $u_r(r,t)$.  The principal stretches can
then be defined as
$\beta_r = 1 + \pdf{u_r}{r}$,
$\beta_\theta = 1 + u_r / r$, and 
$\beta_z(t) = h(t) / H$.
The referential fluid fraction then takes the form
$\phi_w(r,t)$ and can be obtained from $\phi_w = \phi_0 + \beta_r \beta_\theta \beta_z - 1$.  Similarly, we let
the fluid pressure be given by $p(r,t)$.  
Under these assumptions, the balance laws representing
mass and momentum conservation can be reduced to (see Appendix \ref{app:simplification} for details)
\subeq{
\label{eqn:red}
\begin{align}
\frac{r}{2}\pd{}{t}\left(\beta_\theta^2 \beta_z\right) &= \frac{k(J) J}{\beta_r^2}\pd{p}{r}, \label{eqn:p} \\
  \frac{r}{2}\pd{}{t}\left(\beta_\theta^2 \beta_z\right) &=
  \frac{k(J)}{\beta_r}\left[  \pd{S_r^e}{r} + \frac{S_r^e - S_\theta^e}{r} - \beta_\theta \beta_z \pd{\Pi}{r}\right].
  \label{eqn:u}
\end{align}
}
The mass balance \eqref{eqn:p} determines the fluid pressure $p$ and must be solved with the 
boundary condition $p(R,t) = 0$, reflecting the continuity of
fluid stress at the free boundary of the sample.
The momentum balance \eqref{eqn:u} determines the radial
displacement $u_r$ and must be solved with the boundary
conditions $u_r(0,t) = 0$ and $S_r(R,t) = 0$.  The former
represents symmetry at the origin and the latter represents
continuity of total stress at the free boundary.  
The initial conditions for \eqref{eqn:u} are given by
$u_r(r,0) = 0$ and $\beta_z(0) = 1$.

In a force-controlled experiment, the axial stretch $\beta_z$ is
determined from an axial force balance given by
\begin{align}
  F = 2 \pi \int_{0}^{R} (S_z^e - p \beta_r \beta_\theta) r\,\d r,
  \label{eqn:F}
\end{align}
where $F$ is the applied force.  We use the convention
that $F < 0$ corresponds to axial compression.  In this
case, \eqref{eqn:red}--\eqref{eqn:F} must be simultaneously
solved.  Once the displacement is found, the radial and
circumferential stretches, along with the porosity, 
can be computed.  

In a displacement-controlled experiment, the axial stretch
$\beta_z(t)$ is prescribed.   In this case, the equation for the displacement \eqref{eqn:u} decouples and can be solved independently from the rest of the equations.
Once the displacement is known, the pressure $p$ can be computed
by integrating \eqref{eqn:p}. 
The force on the platten can be obtained by evaluating
\eqref{eqn:F}.

\subsection{Equilibrium response}
\label{sec:equilibrium}

The equilibrium response of the material to a load 
is found by setting the time derivatives
to zero and seeking a homogeneous deformation with
$\beta_\theta = \beta_r$.  The balance of fluid pressure
results in $p = 0$ everywhere.  The referential water fraction
can be written in terms of the stretches $\beta_r$ and
$\beta_z$ as $\phi_w = \beta_r^2 \beta_z - 1 + \phi_0$. 
For displacement-controlled loading where
$\beta_z$ is known, 
the radial stretch $\beta_r$ is found by solving
$S^e_r - \beta_r \beta_z \Pi(\phi_w) = 0$.
For force-controlled loading,
the radial and axial stretches must be found by simultaneously solving
$S^e_r - \beta_r \beta_z \Pi(\phi_w) = 0$ and
$F = \pi R^2 [S^e_z - \beta_r^2 \Pi(\phi_w)]$.


\subsection{Instantaneous response}
\label{sec:instant}

The instantaneous response  
can be derived by assuming
that no fluid is expelled ($\phi_w = \phi_0$) on short time scales
and that the
sample remains homogeneous with
$\beta_\theta = \beta_r = \beta_z^{-1/2}$. The fluid in the gel
becomes pressurised with a uniform pressure $P > 0$ that
exceeds the pressure of the surrounding fluid.
The balance
of total stress at the side of the sample leads to
$P = \beta_z^{-1/2} S_r^e - \Pi(\phi_0)$. If $\beta_z$ and hence $\beta_r$ and
$\beta_\theta$ are known due to displacement-controlled loading, then the fluid pressure $P$ can be calculated.  However,
if $\beta_z$ is unknown due to the loading being controlled by force, then the force balance
$F = \pi R^2(S_z^e - \beta_z^{-3/2} S_r^e)$
must first be solved. 

\subsection{Numerical solution}

The open-source Python package {\tt ucompress.py} has been developed 
to provide high-level user-friendly code for simulating the unconfined compression
of nonlinear poroelastic materials using the models developed here. 
The governing 
equations are discretised in space using a Chebyshev spectral method and in time using a fully implicit Euler scheme with an analytical Jacobian matrix.
The package, along with tutorials, is available at \url{https://github.com/hennessymatt/ucompress.py}.

\subsection{Comparison with experiments}
\label{sec:exp}

The microstructural model is compared with
data produced from unconfined compression
experiments carried out by Moore~\etal\cite{moore2023}.  
Recall that the parameter values associated with these materials
are provided in Sec.~\ref{sec:params}.  
Using the initial porosity of the material after
hydration ($\phi_0 = 0.82$), the Flory interaction
parameter has been calculated by solving the
hydration equations presented in Sec.~\ref{sec:hydration}; we find that $\chi = 0.57$ for all samples. 

\begin{figure}
\centering
\subfigure[$\Phi_f = 0.26$]{\includegraphics[width=0.45\textwidth]{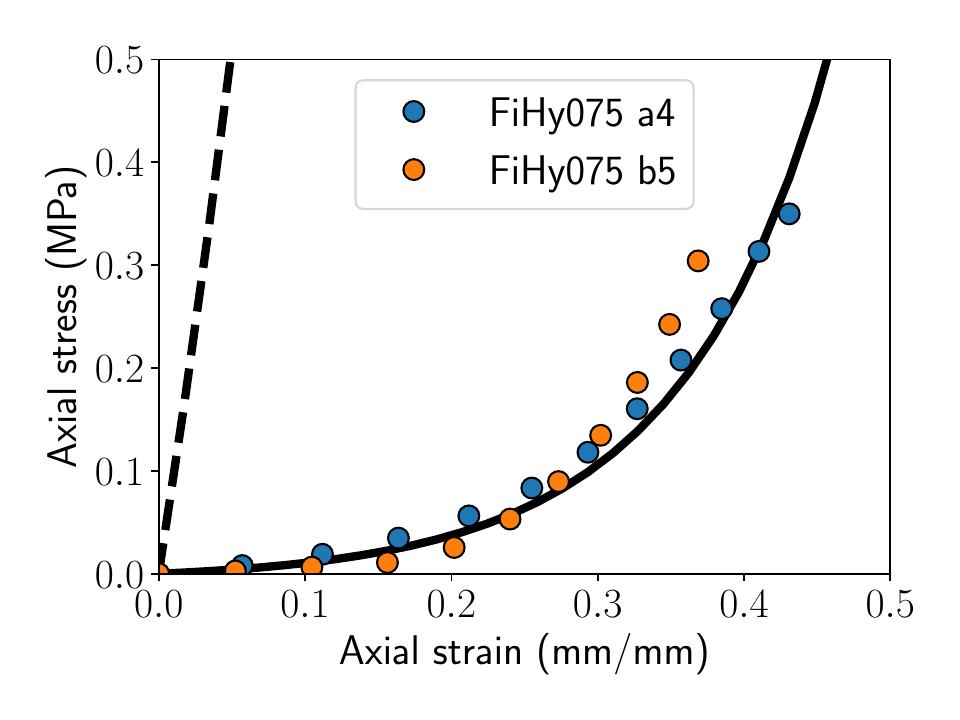}}
\subfigure[$\Phi_f = 0.42$]{\includegraphics[width=0.45\textwidth]{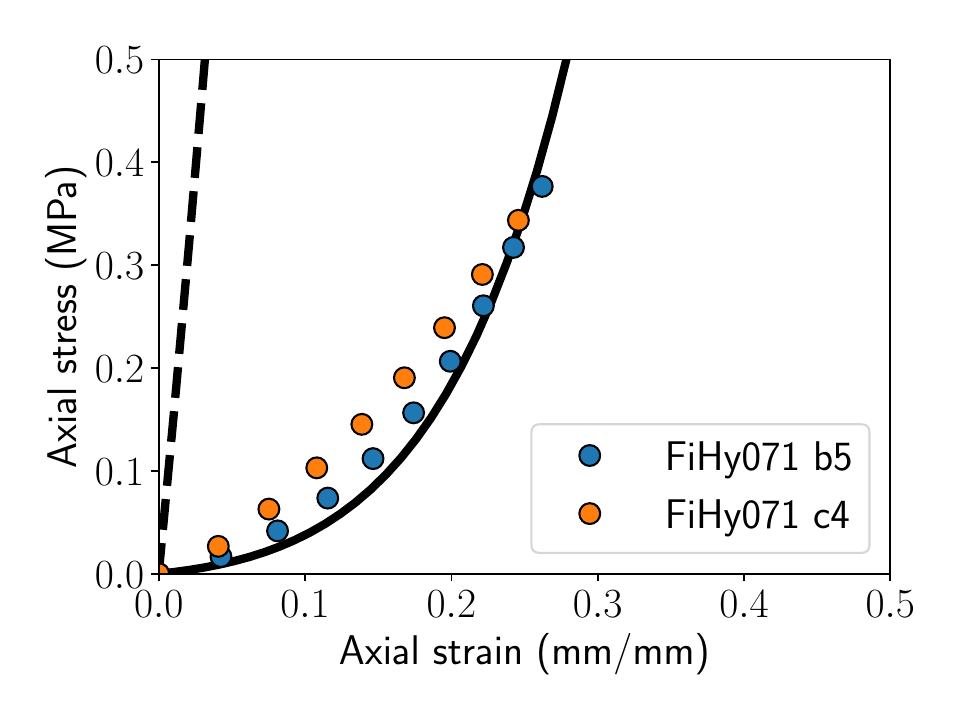}}
\\
\subfigure[$\Phi_f = 0.51$]{\includegraphics[width=0.45\textwidth]{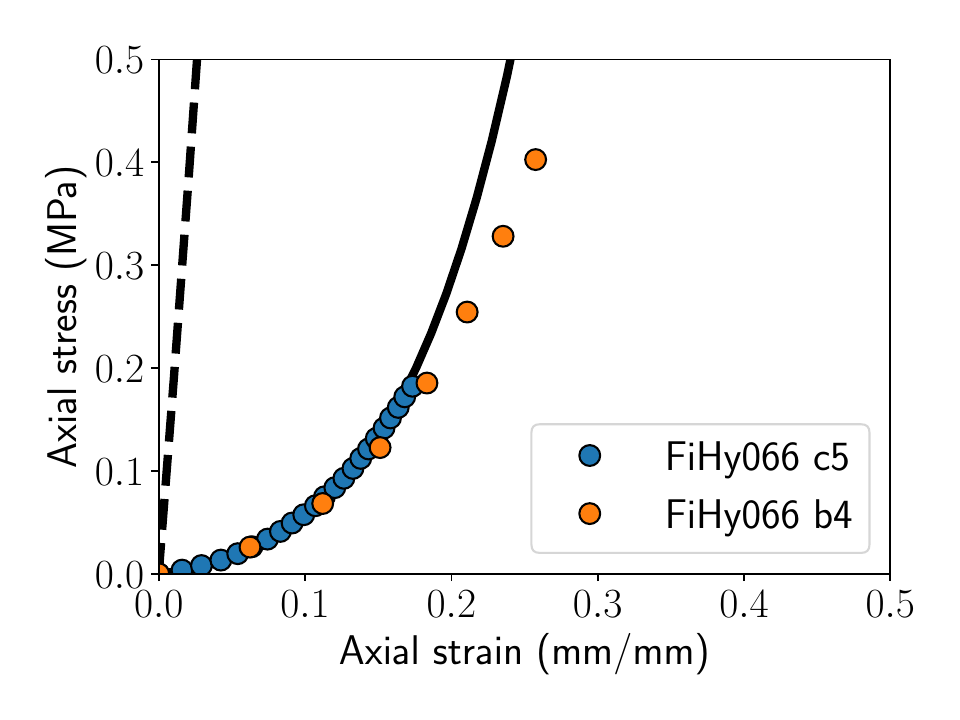}}
\subfigure[$\Phi_f = 0.77$]{\includegraphics[width=0.45\textwidth]{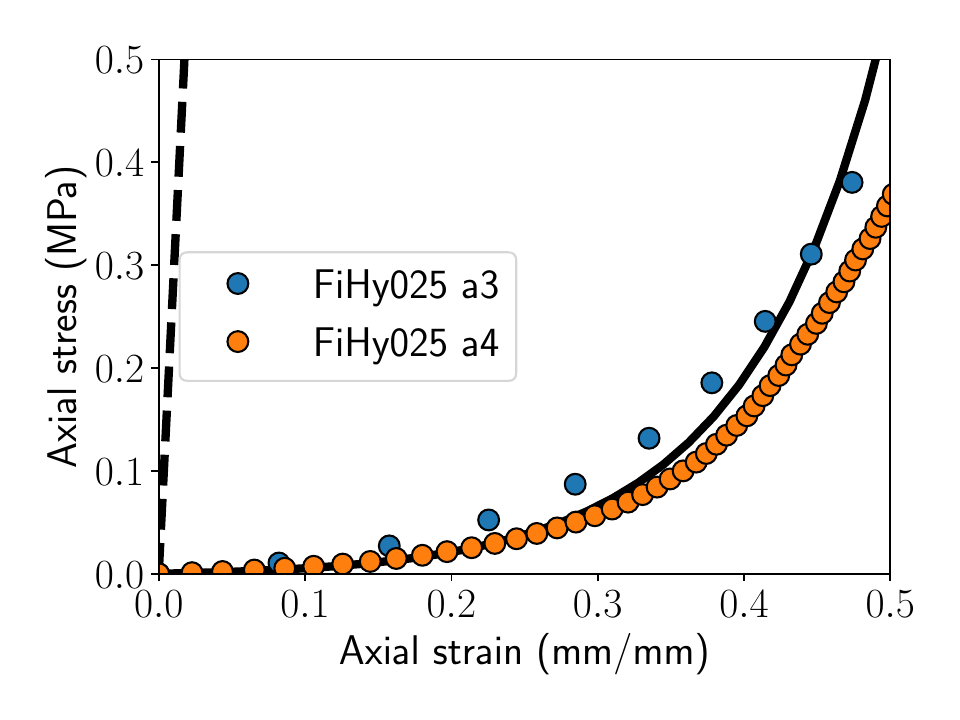}}
\caption{Comparison of the model (lines) with experimental (symbols) stress-strain
data.  Different panels correspond to
materials with different nominal fibre 
fractions $\Phi_f$.  Solid lines correspond to models that account for fibre recruitment; dashed lines correspond to models without recruitment.  The maximum recruitment stretch $\lambda_m$
for each panel is different; see text for details.
Positive axial strains in these figures correspond to compressions rather than stretches.}
\label{fig:stress_strain}
\end{figure}

The microstructural model is first compared with stress-strain data produced by subjecting cylindrical
FiHy samples with different nominal fibre fractions
$\Phi_f$ to compressive axial loads. For each
fibre fraction, $\Phi_f$, two different samples
were fabricated and tested, resulting in 
two sets of stress-strain
data for a given fibre fraction, $\Phi_f$ (Fig.~\ref{fig:stress_strain}).
The short duration of the loading,
which was on the order of 10 seconds, is
much smaller than the poroelastic time scale
over which fluid is squeezed out of the
sample (on the order of 1000 seconds). 
Hence, the stress-strain data is assumed
to represent the instantaneous response 
of the material.  

The stress-strain
data can be modelled by considering
a range of compressive axial stretches $\beta_z\le1$, corresponding
to compressive axial strains of $\epsilon_z = 1 - \beta_z$, and computing the axial stress, $S_z$,  following the procedure outlined
in Sec.~\ref{sec:instant}.  If all
fibres in the network are assumed to
be mechanically active (no recruitment; $\lambda_m = 1$),
then there are no free parameters in the model. However, the predicted stresses (the dashed lines in Fig.~\ref{fig:stress_strain}) greatly exceed those measured in 
experiments.
This is a strong indication that, despite
the materials being pre-loaded due to
hydration, some mechanically
inactive fibres remain in the network.
By accounting for inactive fibres
in the model and fitting the maximum
recruitment stretch $\lambda_m$ to 
experimental data, excellent agreement
is obtained.  When fitting $\lambda_m$,
only one data set for a given 
fibre fraction is used, yet the
model is able to capture both data sets reasonably well.
The value of $\lambda_m$ obtained
by fitting varies with
$\Phi_f$.  We find that for 
$\Phi_f = 0.26$, $0.42$, $0.51$
and $0.77$, the best-fitting maximum recruitment stretch
is $\lambda_m = 3.0$, $1.8$, $1.7$, and
$3.9$, respectively.  The non-monotonic
relationship between the maximum recruitment
stretch and fibre fraction is evident in
Fig.~\ref{fig:stress_strain}; the samples
with $\Phi_f = 0.42$ and $0.51$ exhibit
a much more rapid increase in stress with strain, characteristic of a stiffer material caused by more fibre engagement.  The variability in the values
of $\lambda_m$ obtained from fitting is likely due to material inhomogeneity caused by the electrohydrodynamic fabrication
process.

For the remainder of this section,
we focus on FiHy materials that have a nominal fibre fraction $\Phi_f = 0.42$, as these 
were subjected to additional experimental testing. By taking the maximum recruitment stretch to be $\lambda_m = 1.8$, in line
with Fig.~\ref{fig:stress_strain}~(b), the
mechanical properties of the material are fully
characterised.  The only remaining fitting parameters are related to the permeability and only
influence the time-dependent solutions of the
model.  After hydration, the radius and
height of the sample were $R = 5.165$~mm 
and $h = 1.740$~mm, respectively.

\begin{figure}
\centering
\subfigure[Instant response]{\includegraphics[width=0.48\textwidth]{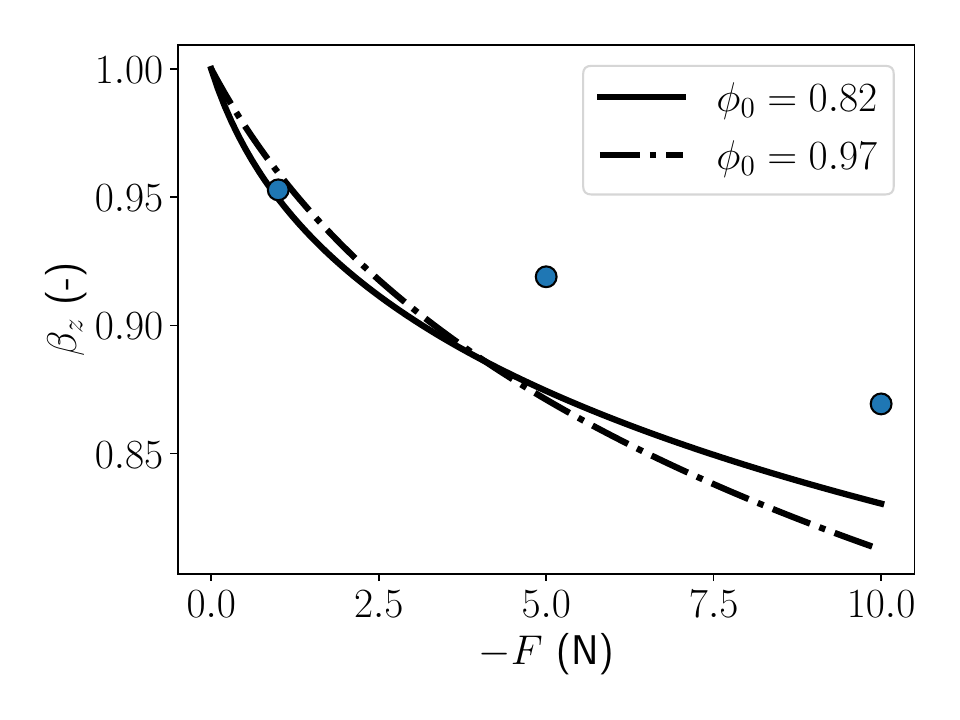}}
\subfigure[Equilibrium response]{\includegraphics[width=0.48\textwidth]{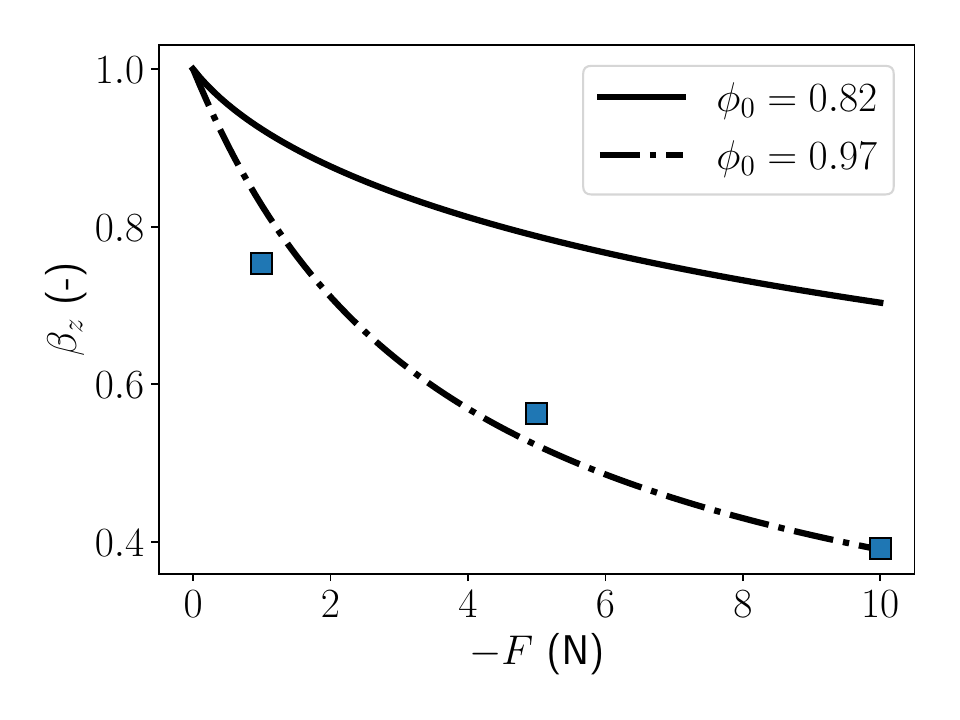}}
\caption{Comparison of model predictions (lines) and experimental measurements (symbols) of the (a) instantaneous and (b) equilibrium response of a fibre-reinforced
hydrogel to an applied force. We plot the axial stretch $\beta_z$ as a function of the applied force $F$. Two values
of the initial porosity $\phi_0$ are
considered.}
\label{fig:instant_steady}
\end{figure}

The instantaneous and equilibrium response
of a FiHy material to different compressive loads was measured.  Specifically, the axial
stretch $\beta_z$ was measured as a function of the applied compressive force $F$.  The data is shown in Fig.~\ref{fig:instant_steady}.  The model overestimates the magnitude of the instantaneous deformation, but underestimates the equilibrium deformation.  The latter suggests there is
a long-term softening mechanism at play.  One explanation for this could
be due to an underestimation of the initial degree of hydration.
From the equilibrium analysis in Sec.~\ref{sec:equilibrium}, we see
that the applied axial load is balanced by the axial elastic stress 
and the osmotic stress.  Since the osmotic stress decreases
with fluid volume fraction, materials
with increased hydration will experience
greater axial deformation upon application
of a force.  Indeed, by setting $\chi = 0.41$
to increase the fluid fraction of the initial hydrated state
to $\phi_0 = 0.97$, keeping all other parameters
fixed, a much better agreement
between the theoretical and experimental
equilibrium response is obtained (the dashed-dotted lines in Fig.~\ref{fig:instant_steady}~(b)).  A second possible mechanism for
long-term material softening is solid-phase
viscoelasticity, which is expected for fibre-reinforced hydrogels~\cite{strange2014} but is not considered in the model here.

\begin{figure}
\centering
\subfigure[$\phi_0 = 0.86$]{\includegraphics[width=0.48\textwidth]{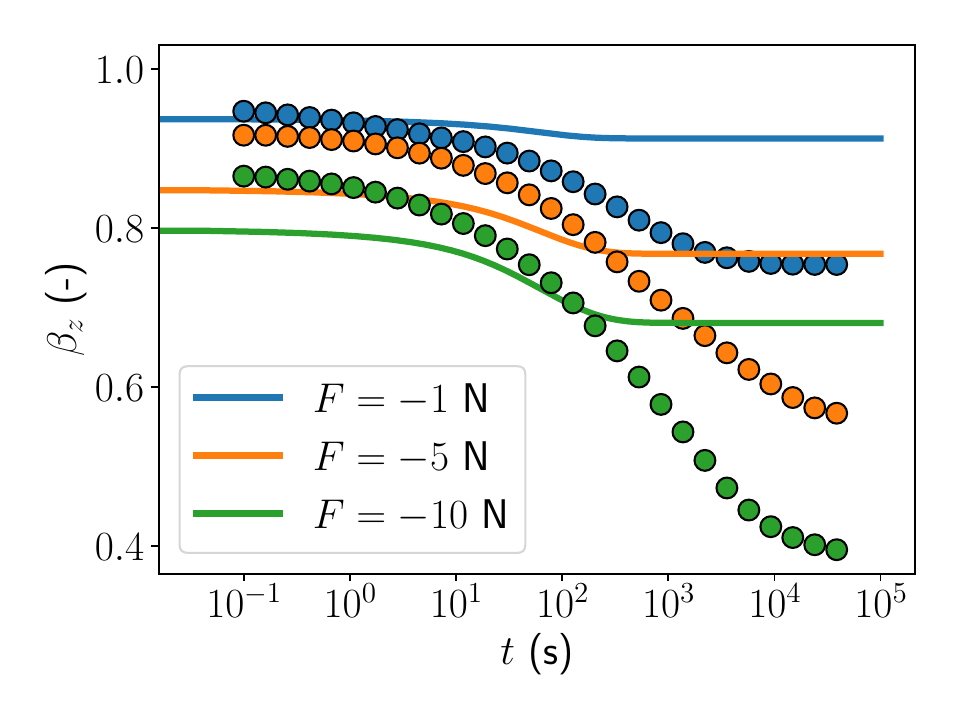}}
\subfigure[$\phi_0 = 0.97$]{\includegraphics[width=0.48\textwidth]{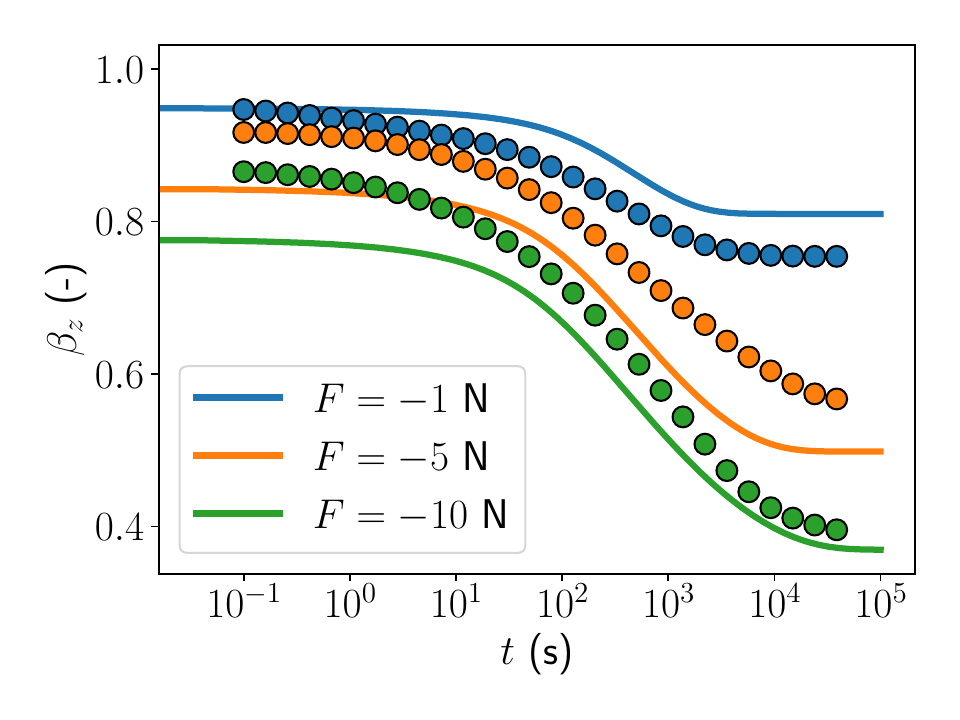}}
\caption{Comparison of model predictions (lines) and experimental measurements (symbols) of the transient response of a fibre-reinforced hydrogel to different applied forces $F$.  The time dependence of the axial stretch $\beta_z$ is shown
for two different values of the initial
porosity $\phi_0$.}
\label{fig:transient}
\end{figure}

The time-dependent response of a FiHy material
to an axial compressive force was also
studied by Moore \etal\cite{moore2023}.  In particular, the axial stretch $\beta_z$ was measured as a function of time for three applied
loads, $F$.  For small times, there are plateaus in the data that coincide with the instantaneous
response of the material described in Sec.~\ref{sec:instant} (see Fig.~\ref{fig:transient}~(a)).  However,
a plateau in the data for large times is
only seen when the applied load is small 
($F = -1$~N), providing further evidence
that the material exhibits slow
viscoelastic creep.  

To simulate the time-dependent
material response, we set the parameters in the Mow--Holmes permeability law
\eqref{eqn:perm} to be $k_0 = 2 \times 10^{-13}$ m$^2$ Pa$^{-1}$ s$^{-1}$, $\gamma = 4$, and $M = 2$.  The model predictions (the solid lines in Fig.~\ref{fig:transient}~(a)) capture the qualitative features of the
data.  However, large quantitative discrepancies occur; this is mainly
due to the inability of the model to
capture the instantaneous and equilibrium
responses of the material shown in
Fig.~\ref{fig:instant_steady}.  Changes to
the permeability-related parameters do
not strongly impact the agreement between
model and experiment, as variations in
$k_0$ only horizontally shift the model
curves, whereas variations
in $\alpha$ and $M$ alter the transition
between the instantaneous and
equilibrium plateaus.  However,
by increasing the initial porosity to $\phi_0 = 0.97$, the time-dependent response predicted
by the model is in much better agreement 
with the experimental data (Fig.~\ref{fig:transient}~(b)). These
results underscore the importance of
accurately modelling the instantaneous
and equilibrium response of the material, which may require incorporating additional
physics into the model.
Nevertheless, with only a minimal number
of fitting parameters, the microstructural
model is able to capture all of the experimental data qualitatively, and, in
some cases, provide excellent quantitative agreement as well.

\section{Design of materials for artificial cartilage}
\label{sec:design}

Having validated our model, we now showcase how it can be used 
to guide the design of fibre-reinforced hydrogels used as artificial cartilage.  One of the most remarkable features of 
cartilage is that it is able to bear compressive loads that 
greatly exceed its equilibrium compressive 
modulus~\cite{moore2023}. This
load-bearing capacity occurs through the ability of cartilage
to transfer load from its porous, solid matrix to the fluid
residing in the pore space.  Unconfined compression experiments
have shown that the fraction of load supported by the
interstitial fluid 
(also called the fluid load fraction) can be as high as
79\% in human cartilage, and 94\% in bovine cartilage~\cite{park2003}.  Thus, any synthetic material
used to replace cartilage should achieve similar
fluid load fractions.

The goal of this section is to use our model to predict,
understand, and optimise the fluid load fraction in
transversely isotropic fibre-reinforced hydrogels during
unconfined compression.
We consider cylindrical
gelatin-PCL composites that are based on the FiHy materials
produced by Moore \etal\cite{moore2023}.  Therefore, we take
$E_m = 4$~kPa, $\nu_m= 0$, and $E_f = 400$~MPa.
The maximum recruitment stretch, $\lambda_m$, and the nominal fibre fraction, $\Phi_f$, are varied.
The radius and height of the hydrated sample are taken to
be $R = 5.165$~mm and $h_0 = 1.740$~mm, respectively; the Flory interaction parameter
is set to $\chi = 0.57$.  For simplicity, the permeability is taken to 
be a constant given by $k = 2 \times 10^{-13}$ m$^2$ Pa$^{-1}$ s$^{-1}$;
however, using the Holmes--Mow permeability law given by
\eqref{eqn:perm} leads to similar results.
The axial compressive force is set to $F = -10$~N.  
We assume that the simplifications in Sec.~\ref{sec:unconfined}
still apply; namely, the material can slip along the plattens so that
the cylindrical shape is preserved during compression.  In this case, the fluid load fraction can be defined as
$-\bar{p} A / F$, where
\begin{align}
\bar{p} = \frac{2 \pi}{A} \int_{0}^{R} p \beta_r \beta_\theta r \,\d r,
\label{eqn:mean_p}
\end{align}
is the radially averaged fluid pressure, $F$ is the applied compressive force, and $A$ is the area in the current configuration of the upper and lower cylindrical surfaces.
If the pressure and stretches are uniform, with $\beta_\theta = \beta_r$, then the fluid load fraction reduces to 
$-p A / F$ where $A = \pi R^2 \beta_r^2$.

\subsection{Numerical simulations of the fluid load fraction}

\begin{figure}
\centering
\includegraphics[width=0.85\textwidth]{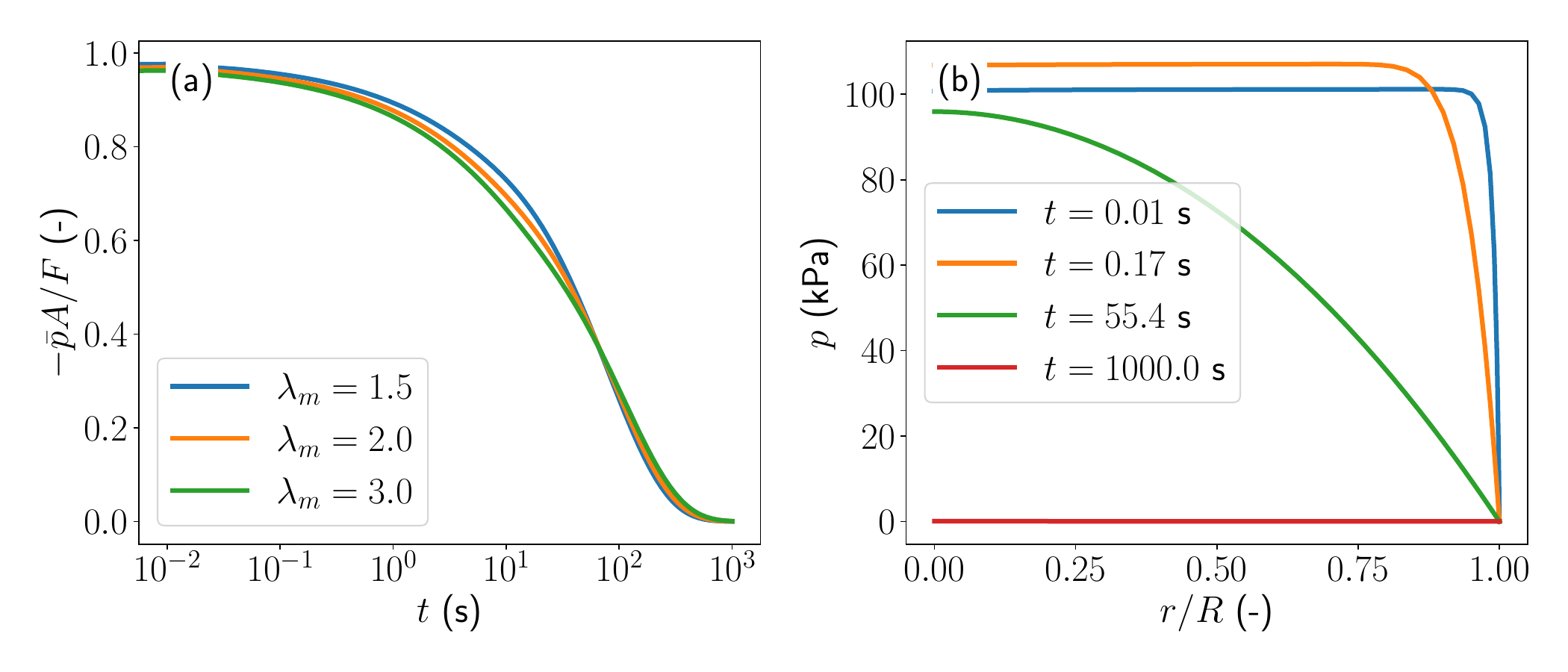}
\caption{(a) Fluid load fraction as a function of time for different
maximum recruitment stretches.  (b) Spatial profiles of the
fluid pressure at different times when the
maximum recruitment stretch is $\lambda_m = 1.5$.}
\label{fig:transient_flf}
\end{figure}

The evolution of the fluid load fraction was numerically calculated
for a fixed nominal fibre fraction of $\Phi_f = 0.4$ and a maximum
recruitment stretch of $\lambda_m = 1.5, 2.0, 3.0$.  When
a load is applied, the fluid load fraction instantaneously achieves
its peak value.  Afterwards, it monotonically decays
to zero due to poroelastic relaxation; see Fig.~\ref{fig:transient_flf}~(a).
More specifically, the application of a force increases the
fluid pressure above zero, creating a pressure gradient that
pushes fluid out of the composite and into the surrounding
environment until equilibrium is again established; 
see Fig.~\ref{fig:transient_flf}~(b).

Numerical calculations show that, as the maximum recruitment stretch
increases from $\lambda_m = 1.5$, corresponding
to fibrous networks with more initial slack, 
the peak fluid  load fraction decreases and the duration over which the peak fluid load fraction 
persists is shortened (Fig.~\ref{fig:transient_flf}~(a)).  Thus,
removing slack from the fibre network can have a two-fold positive impact on
the performance of the material.  However, removing slack does accelerate
the overall relaxation of the fluid load fraction, causing the new
equilibrium state to be reached sooner.  This behaviour is expected because
the poroelastic relaxation time is inversely proportional to the
radial Young's modulus, which will be larger for materials with fibre networks that are more engaged.

\subsection{Insights from the peak fluid load fraction}

For the remainder of this section, we focus on the peak fluid
load fraction, which can be obtained by solving for the instantaneous
response of a sample when a compressive load is applied.
From Sec.~\ref{sec:instant}, the instantaneous fluid
pressure is given by $p = P = \beta_z^{-1/2} S_r^e - \Pi$
and is therefore controlled by two competing mechanisms.  
Firstly, the radial elastic stresses act to pressurise the fluid. Secondly, the tensile osmotic stress, which resists axial compression,
acts to depressurise the fluid.  Thus, maximising the peak
fluid load fraction amounts to decreasing the osmotic stress
and increasing the radial elastic stress.

\begin{figure}
\centering
\includegraphics[width=0.45\textwidth]{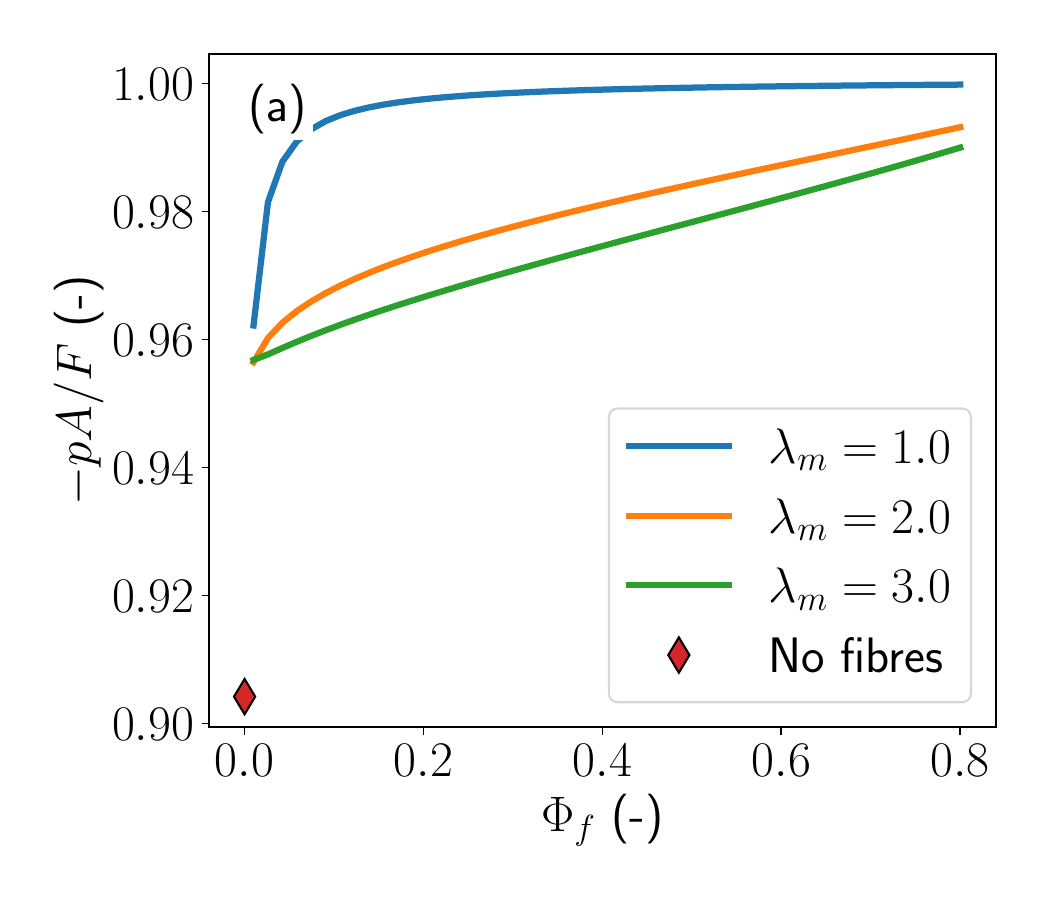}
\includegraphics[width=0.45\textwidth]{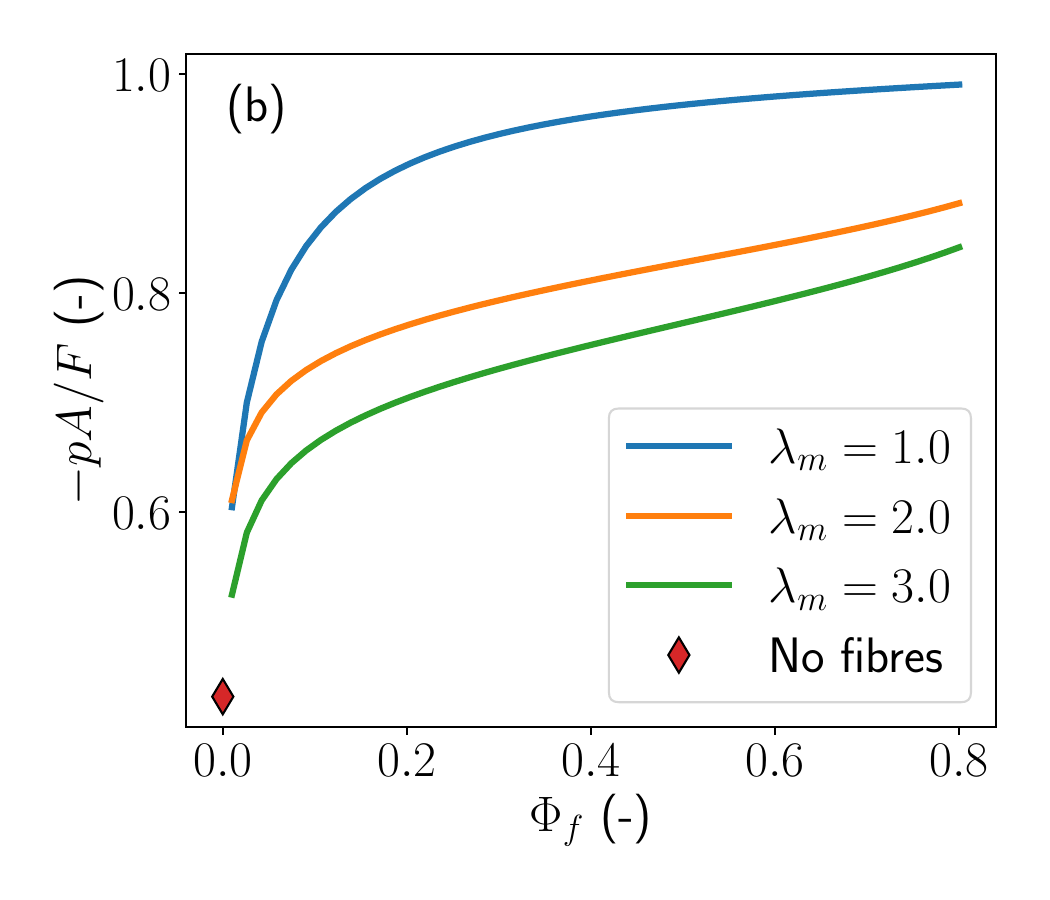}
\caption{Fraction of the load supported by the fluid
pressure (fluid load fraction) 
as a function of the nominal fibre
fraction $\Phi_f$ for various maximum recruitment
stretches $\lambda_m$. The Young's modulus of the hydrogel
matrix is (a) 4~kPa and (b) 400~kPa.
The curves are computed by
solving for the instantaneous response of the sample
to a 10~N compressive axial force.  The quantity
$A = \pi R^2 \beta_r^2$ is the current area of the 
cylindrical face on which the axial force is applied.}
\label{fig:peak_flf}
\end{figure}

During the instantaneous response, the amount of fluid
contained within the material is fixed.  As a result, the 
osmotic stress
is equal to that of the initially hydrated state.  The osmotic 
stress
of the initially hydrated state must balance the axial elastic
stress (see Sec.~\ref{sec:hydration}). Hence, the osmotic stress
scales roughly
with the Young's modulus of the hydrogel matrix and
the nominal volume fraction of the matrix, $1 - \Phi_f$,
so that $\Pi \sim (1 - \Phi_f) E_m$.  Increasing the
nominal fibre fraction, $\Phi_f$, is expected to decrease the osmotic stress and increase
the peak fluid load fraction, as confirmed by numerical results,
which show the fluid load fraction exceeding 95\%;
see Fig.~\ref{fig:peak_flf}~(a).  The large fluid load
fraction across the entire range of fibre
fractions suggests that osmotic stresses only support
a small fraction of the load.  Indeed, the above scaling analysis
shows that the osmotic stress is roughly $\Pi \sim 1$~kPa in
magnitude,
whereas the compressive stress induced by the load is on the 
order of $F / (\pi R^2) \sim 100$~kPa.  Increasing
the hydrogel modulus by a factor of 100 to $E_m = 400$~kPa
results in greatly increased
osmotic stresses and markedly lower peak fluid load fractions 
(Fig.~\ref{fig:peak_flf}~(b)).

The radial elastic stress also plays a key role in controlling
the peak fluid load fraction, and is dependent on both
the Young's modulus of the fibres and the maximum 
recruitment stretch, $\lambda_m$.  The impact of the
fibre modulus on fluid load fraction has been studied by
Tan \etal\cite{tan2023}; thus, we focus on the role of
fibre engagement.  As the maximum recruitment stretch
increases, there are more inactive fibres in the network and the
radial stiffness decreases.  As a result, the fluid becomes less pressurised due to the applied force
and the fluid load fraction decreases, as can be seen in
Fig.~\ref{fig:peak_flf}.  As an extreme case, we have
computed the instantaneous response of a fibre-free network.  The
lack of any radial reinforcement 
leads to a marked decrease in the
fluid load fraction;
see the red diamonds in Fig.~\ref{fig:peak_flf}.  Thus,
radial reinforcement plays an important role in generating interstitial fluid pressure.

Moore \etal\cite{moore2023} inferred the peak fluid load fraction of FiHy
samples from unconfined compression experiments.  
They found that the fluid load fraction varied non-monotonically with
the nominal fibre fraction.  Generally, the fluid load fraction was
maximised when the (target) ratio of gelatin to PCL was 2:1 by mass. 
Smaller or larger proportions of gelatin led to a decrease in the fluid
load fraction.
The results in 
Fig.~\ref{fig:peak_flf} indicate
that the fluid load fraction is sensitive to the amount of slack in the
fibre network, with the fluid load fraction increasing with decreasing
slack (or values of $\lambda_m$).  When fitting the stress-strain data in Fig.~\ref{fig:stress_strain}, we found that materials with intermediate fibre fractions had the least amount of slack
and, therefore, would be expected to 
have the greatest fluid load fractions, which matches with the 
findings  of Moore \emph{et al}. Thus, the maximum in the fluid load
fraction observed by Moore \emph{et al}. is likely due to variation
in the amount of slack in the fibre network across the samples.

\subsection{Design implications}

Our analysis of the peak fluid load fraction suggests that the
load-bearing capacity of the fluid can be enhanced by 
decreasing the ratio of the axial stiffness to the radial stiffness.
Decreasing the axial stiffness by using a softer hydrogel matrix or
a greater volume fraction of fibres will decrease the osmotic stresses,
allowing the fluid to bear more of the applied load.  Increasing the
radial stiffness by introducing a reinforcing fibrous network,
decreasing the slack in the network, or increasing the fibre stiffness
will increase the pressurisation of the interstitial fluid. 

The degree of swelling can also play an important role in controlling
the fluid load fraction.  We find that increasing the degree of swelling
(decreasing $\chi$ but keeping other parameters fixed) reduces
the fluid load fraction by increasing the osmotic
stress.  In theory, the volumetric expansion
of the material due to hydration could reduce the slack in the fibre
network, thereby increasing the fluid load fraction.  However, the
degree of in-plane expansion is set by the ratio of the osmotic stress
to the radial Young's modulus and is likely to be small for strongly
reinforced materials; that is, materials with a large contrast in their radial and
axial Young's moduli.  Therefore, the ability for swelling to increase
the fluid load fraction through fibre engagement could be limited.

\section{Conclusion}
\label{sec:conclusion}

In this paper, we developed a mathematical framework for modelling fibre-reinforced hydrogel composite materials subjected to finite deformations. In particular, we derived a microstructural model of the fibre phase, incorporating the gradual recruitment of fibre segments due to fibre waviness, and upscaled it to a nonlinear elastic framework. This was combined with a poroelastic model of the hydrogel phase, which allowed us to model the swelling of the gel during hydration explicitly, before considering subsequent loading. The model could be equally applied to manufactured biomaterials, or collagen-based biological tissues, such as articular cartilage.

Due to the simplicity of our modelling framework compared to
other multi-scale approaches for fibre-reinforced hydrogels~\cite{castilho2019, chen2020}, 
it is possible to quickly sweep through parameter 
space to determine how changes to the material
properties impact the poromechanical response of
the composite.  This feature allows the model
to serve as a useful tool for guiding the design
of new materials, which we showcased in the context of fabricating artificial cartilage.  Although we considered fibre-reinforced materials
that are transversely isotropic, it is possible to extend the model to
more general orthotropic materials, such as those used 
by Visser \etal\cite{visser2015} and Bas \etal\cite{bas2015},
by using an appropriate probability density function for the
orientation of fibres in \eqref{WTheta} or \eqref{Wtot}. 
Thus, our framework opens the door to optimising the fibre
network geometry for specific applications.

The model implicitly assumes that, whilst their slack is being stretched out, the fibres are able to slip frictionlessly through the hydrogel without mechanical interaction, whereas once fully recruited, they move identically to the matrix. This is a limitation that should be addressed in future work. Nevertheless, the model is able to capture several key features of fibre-reinforced hydrogels that have been observed experimentally. In particular, the gradual recruitment of fibres allows accurate quantitative agreement with the nonlinear, J-shaped loading curves seen in Fig.~\ref{fig:stress_strain}, as well as qualitative agreement with the time-dependent behaviour displayed in Fig.~\ref{fig:transient}. To improve the quantitative agreement with the transient behaviour, it will likely be necessary to incorporate constitutive viscoelasticity into the fibres themselves. Several non-linear viscoelastic phenomena can be explained by the gradual recruitment of viscoelastic fibres~\cite{shearer2020recruitment}.

Finally, the fibres in our model are assumed to be infinitely flexible, with no bending stiffness. A potential further improvement could be to incorporate the influence of the fibres' aspect ratio on their bending stiffness into the model and investigate the corresponding mechanical effects on the network. However, it may be challenging to scale such a model up analytically to the macroscale as we have done here; thus, as always, there is a balance to be struck between the accuracy and the tractability of future models.


\appendix

\section{Analytical derivation of the strain energy function}
\label{app}

\begin{figure}
\centering   \includegraphics[width=\textwidth]{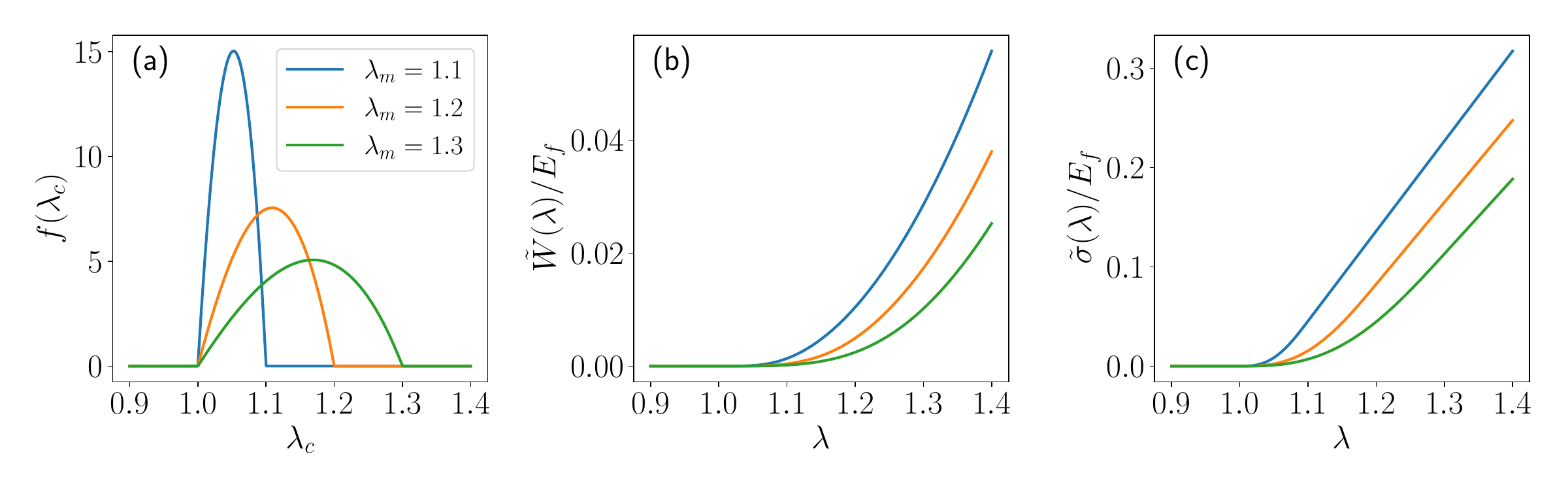}
\caption{(a) The quartic recruitment stretch distribution function, $f(\lambda_c)$, given by \eqref{app:quartic}. (b) The average fibre segment strain energy density, $\tilde{W}(\lambda)$.  (c) The average fibre segment stress, $\tilde{\sigma}(\lambda)$.}
\label{fig:quartic}
\end{figure}

In this appendix, we define a probability density function, $f(\lambda_c)$, such that the double integral
\eqref{Wtot} can be evaluated analytically. 
When a standard probability density function (e.g.\ exponential,
gamma, etc) is used, the $\lambda/\lambda_c$ term in equation \eqref{W(lambda)} leads to $\lambda\log\lambda$ terms, which cannot be integrated analytically in equation \eqref{Wtot} once $\lambda=\sqrt{\lambda_1^2\cos^2\Theta+\lambda_2^2\sin^2\Theta}$ is substituted in. Distribution functions that eliminate these terms admit an analytical solution. For example,
the quartic distribution given by
\begin{align}
    f(\lambda_c)=\begin{cases}
        0, & \lambda_c\le1,\\
        \dfrac{60\lambda_c^2(\lambda_c-1)(\lambda_c-\lambda_m)}{3-5\lambda_m+5\lambda_m^4-3\lambda_m^5}, & 1\le\lambda_c\le\lambda_m,\\
        0, & \lambda_c\ge\lambda_m,
    \end{cases}
    \label{app:quartic}
\end{align}
where $\lambda_m$ is the maximum recruitment stretch, has the required properties. It is a probability density function as it
satisfies $\int_0^\infty f(\lambda_c)\,\d\lambda_c=1$. It also ensures that no fibres are pre-loaded in the reference configuration as the minimum recruitment stretch is $\lambda_c=1$. 
This quartic distribution is plotted in Figure \ref{fig:quartic}~(a).

Upon evaluating equation \eqref{W(lambda)} using the quartic 
distribution \eqref{app:quartic}, we find that the
stretch-averaged strain energy of the fibres is
\begin{align}
   \tilde{W}_f(\lambda)=\begin{cases}
   0, & \lambda\le1, \\
   \dfrac{E_f(\lambda-1)^4(5\lambda_m-3-2\lambda)}{2(\lambda_m-1)^3(3+4\lambda_m+3\lambda_m^2)}, & 1\le\lambda\le\lambda_m,\\
       \dfrac{E_f}{2}+\dfrac{5E_f\lambda(\lambda-\lambda_m-1)}{(3+4\lambda_m+3\lambda_m^2)}, & \lambda\ge\lambda_m,
   \end{cases} 
\end{align}
which gives rise to an average fibre segment stress ($\tilde{\sigma}_f = \d \tilde{W}_f / \d \lambda$) of
\begin{align}
   \tilde{\sigma}_f(\lambda)=\begin{cases}
   0, & \lambda\le1, \\
   \dfrac{5E_f(\lambda-1)^3(2\lambda_m-1-\lambda)}{(\lambda_m-1)^3(3+4\lambda_m+3\lambda_m^2)}, & 1\le\lambda\le\lambda_m,\\
       \dfrac{5E_f(2\lambda-\lambda_m-1)}{(3+4\lambda_m+3\lambda_m^2)}, & \lambda\ge\lambda_m.
   \end{cases} 
\end{align}
These functions are plotted in Fig.~\ref{fig:quartic} (b) and (c).
Upon evaluating the double integral in equation \eqref{Wtot}, we obtain the mean strain energy of the fibre network,
\begin{align}
   W_f=\begin{cases}
   0, & \dfrac{2\lambda_1}{\pi}\mathcal{E}\left(1-\dfrac{\lambda_2^2}{\lambda_1^2}\right)\le1, \\
   w_1(\lambda_1,\lambda_2,\lambda_m), & 1\le\dfrac{2\lambda_1}{\pi}\mathcal{E}\left(1-\dfrac{\lambda_2^2}{\lambda_1^2}\right)\le\lambda_m,\\
   w_2(\lambda_1,\lambda_2,\lambda_m), & \dfrac{2\lambda_1}{\pi}\mathcal{E}\left(1-\dfrac{\lambda_2^2}{\lambda_1^2}\right)\ge\lambda_m,
   \end{cases} 
\end{align}
where
\begin{multline}
w_1(\lambda_1,\lambda_2,\lambda_m)=\dfrac{E_f}{240\pi(\lambda_m-1)^2(3+4\lambda_m+3\lambda_m^2)}\Biggl(15\pi\Bigl(5(1+\lambda_m(3(\lambda_1^4+\lambda_2^4)+2\lambda_1^2\lambda_2^2)-\\
40(1-3\lambda_m)(\lambda_1^2+\lambda_2^2)-4(6-10\lambda_m)\Bigr)+64(2(\lambda_1^2+\lambda_2^2)+25\lambda_m)\lambda_1\lambda_2^2\mathcal{K}\left(1-\frac{\lambda_2^2}{\lambda_1^2}\right)-\\
\left.32(8(\lambda_1^4+\lambda_2^4)+7\lambda_1^2\lambda_2^2+100(\lambda_1^2+\lambda_2^2)\lambda_m-75(1-2\lambda_m))\lambda_1\mathcal{E}\left(1-\frac{\lambda_2^2}{\lambda_1^2}\right)\right),
\end{multline}
\begin{align}
    w_2(\lambda_1,\lambda_2,\lambda_m)=\frac{E_f}{2}+\frac{E_f}{2\pi(3+4\lambda_m+3\lambda_m^2)}\left(5\pi(\lambda_1^2+\lambda_2^2)-20(1+\lambda_m)\lambda_1\mathcal{E}\left(1-\frac{\lambda_2^2}{\lambda_1^2}\right)\right),
\end{align}
and $\mathcal{K}(\cdot)$ is the complete elliptic integral of the first kind. In terms of the two-dimensional strain invariants, this can be written as
\begin{align}
   W_f=\begin{cases}
   0, & \dfrac{1}{\pi}\sqrt{I_1+\sqrt{I_1^2-4I_2}}\mathcal{E}\left(2\dfrac{\sqrt{I_1^2-4I_2}}{\sqrt{I_1^2-4I_2}+I_1}\right)\le1, \\
   w_1^{2D}(I_1,I_2,\lambda_m), & 1\le\dfrac{1}{\pi}\sqrt{I_1+\sqrt{I_1^2-4I_2}}\mathcal{E}\left(2\dfrac{\sqrt{I_1^2-4I_2}}{\sqrt{I_1^2-4I_2}+I_1}\right)\le\lambda_m,\\
   w_2^{2D}(I_1,I_2,\lambda_m), & \dfrac{1}{\pi}\sqrt{I_1+\sqrt{I_1^2-4I_2}}\mathcal{E}\left(2\dfrac{\sqrt{I_1^2-4I_2}}{\sqrt{I_1^2-4I_2}+I_1}\right)\ge\lambda_m,
   \end{cases} 
\label{WtotI1I2}
\end{align}
where
\begin{multline}
w_1^{2D}(I_1,I_2,\lambda_m)=\dfrac{E_f}{240\pi(\lambda_m-1)^2(3+4\lambda_m+3\lambda_m^2)}\Biggl(15\pi\Bigl(5(1+\lambda_m(3I_1^2-4I_2)-\\
40(1-3\lambda_m)I_1-4(6-10\lambda_m)\Bigr)+\frac{32\sqrt{2}(2I_1+25\lambda_m)I_2}{\sqrt{I_1+\sqrt{I_1^2-4I_2}}}\mathcal{K}\left(2\frac{\sqrt{I_1^2-4I_2}}{\sqrt{I_1^2-4I_2}+I_1}\right)-\\
\left.16(8I_1^2-9I_2+100I_1\lambda_m-75(1-2\lambda_m))\sqrt{I_1+\sqrt{I_1^2-4I_2}}\mathcal{E}\left(2\dfrac{\sqrt{I_1^2-4I_2}}{\sqrt{I_1^2-4I_2}+I_1}{\lambda_1^2}\right)\right),
\end{multline}
\begin{multline}
    w_2^{2D}(I_1,I_2,\lambda_m)=\frac{E_f}{2}+\\
    \frac{E_f}{2\pi(3+4\lambda_m+3\lambda_m^2)}\left(5\pi I_1-20(1+\lambda_m)\sqrt{I_1+\sqrt{I_1^2-4I_2}}\mathcal{E}\left(2\dfrac{\sqrt{I_1^2-4I_2}}{\sqrt{I_1^2-4I_2}+I_1}{\lambda_1^2}\right)\right).
\label{gI1I2}
\end{multline}
The strain energy function \eqref{WtotI1I2} can be expressed in terms of the three-dimensional strain invariants by making the
substitutions given in \eqref{eqn:2D_3D}.

\section{Model derivation using non-equilibrium thermodynamics}
\label{sec:thermo}

The constitutive equations for the model
can be derived using non-equilibrium
thermodynamics.  To begin, the balance
laws for mass, momentum, and volume (incompressibility) are written in terms
of the initial, unhydrated state as follows:
\subeq{
\label{app:balance}
\begin{align}
\dot{\Phi}_w + \nabla_0 \cdot \vec{Q}_0 = 0, \\
\nabla_0 \cdot \tens{S}_0 = 0, \\
J_t = 1 + \Phi_w, \label{app:ic}
\end{align}}
where the dot denotes differentiation
with respect to time, $\nabla_0$ denotes the gradient with
respect to coordinates in the unhydrated
state, $\vec{Q}_0$ is the nominal water flux,
and $\tens{S}_0$ is the nominal stress,
both measured with respect to areas in 
the unhydrated state.  
We also let $\vec{u}_t$ denote the displacement of material
elements relative to their position in the
unhydrated material.  The total
deformation gradient tensor $\tens{F}_t$
is then defined as $\tens{F}_t = \tens{I} + \nabla_0 \vec{u}_t$.
The aim is now to determine expressions for
$\vec{Q}_0$ and $\tens{S}_0$, along with the
chemical potential of water $\mu_w$, which
will be the thermodynamic driving force for
the water flux.

Gurtin~\cite{gurtin1996}
showed that the second law of thermodynamics
can be written as
\begin{align}
\td{}{t}\int_{\mathcal{V}_0} W\,\d V_0 \leq
- \int_{\partial \mathcal{V}_0} \mu_w \vec{Q}_0 \cdot \vec{N}\, \d A_0 +
\int_{\partial \mathcal{V}_0} \tens{S}_0\vec{N}
\cdot \dot{\vec{u}}_t\,\d A_0,
\label{eqn:gurtin_int}
\end{align}
where 
$\d A_0$ and $\d V_0$ are differential
area and volume elements in the unhydrated
state and $\mathcal{V}_0$ and
$\partial \mathcal{V}_0$ denote representative
volume elements and their associated surfaces.
Localising \eqref{eqn:gurtin_int} through
the use of the divergence theorem leads
to
\begin{align}
\dot{W} + \nabla_0 \cdot (\mu_w \vec{Q}_0) - \nabla_0 \cdot (\tens{S}_0^T \cdot \dot{\vec{u}}_t) \leq 0.
\label{app:gurtin}
\end{align}
To incorporate the incompressibility 
condition
\eqref{app:ic} into the energy-imbalance
inequality \eqref{app:gurtin}, 
we differentiate
\eqref{app:ic} with respect to $t$ to find
$J_t \tens{F}_t^{-T}:\dot{\tens{F}}_t - \dot{\Phi}_w = 0$.  Multiplying this result by a
Lagrange multiplier $-p$ and adding
\eqref{app:gurtin} leads to
\begin{align}
\dot{W} + \nabla_0 \cdot (\mu_w \vec{Q}_0) - \nabla_0 \cdot (\tens{S}_0^T \cdot \dot{\vec{u}}) - p(J_t \tens{F}_t^{-T}:\dot{\tens{F}}_t - \dot{\Phi}_w)\leq 0.
\label{app:gurtin_2}
\end{align}
By substituting $W = W_e(\tens{F}_t) + W_\text{mix}(J_t)$ into \eqref{app:gurtin_2},
using the chain rule and the
balance laws \eqref{app:balance}, 
and collecting terms leads to
\begin{align}
\left(\pd{W_e}{\tens{F}_t} + J_t \pd{W_\text{mix}}{J_t} \tens{F}_t^{-T} - p J_t \tens{F}_t^{-T} - \tens{S}_0\right):\dot{\tens{F}}_t +
\left(p - \mu_w\right)\dot{\Phi}_w
+ \nabla_0 \mu_w \cdot \vec{Q}_0 \leq
0.
\label{app:gurtin_3}
\end{align}
Following the procedure established by
Coleman and Noll~\cite{coleman1961}, the
water chemical potential and the nominal
stress tensor can be obtained as
\subeq{
\begin{align}
\mu_w &= p, \\
\tens{S}_0 &= \pd{W_e}{\tens{F}_t} + J_t \pd{W_\text{mix}}{J_t} \tens{F}_t^{-T} - p J_t \tens{F}_t^{-T} = \tens{S}^e_0 - (\Pi + p) J_t \tens{F}_t^{-T}, \label{app:S_0}
\end{align}}
where we have introduced the osmotic stress
$\Pi = -\pdf{W_\text{mix}}{J_t}$ and
the elastic contribution to the stress
tensor  $\tens{S}^e_0 = \pdf{W_e}{\tens{F}_t}$.
The energy-imbalance inequality \eqref{app:gurtin_3}
reduces to $\nabla_0 \mu_w \cdot \vec{Q}_0 \leq 0$.
Thus, by using the equivalance of the chemical potential
and fluid pressure, it follows that the water flux is given by
\begin{align}
\vec{Q}_0 = -\tens{K}_0 \nabla_0 p,
\label{app:Q_0}
\end{align}
where $\tens{K}_0$ is a positive definite 
permeability tensor.
Assuming that the permeability 
in the current configuration is 
isotropic and given by $k$, use
of Nanson's formula shows that the
nominal permeability tensor must be given
by $\tens{K}_0 = k J_t \tens{F}_t^{-1}\tens{F}_t^{-T}$.
The governing equations are now
fully determined.

In Sec.~\ref{sec:balance_laws}, the
governing equations are written in
terms of the initial hydrated state.
Mapping the balance laws and constitutive
equations to the initial hydrated state
from the unhydrated state is straightforward
using Nanson's formula.  In particular,
the nominal and referential 
water flux and stress are related via
\begin{align}
\vec{Q}_w = J_h^{-1} \tens{F}_h  \vec{Q}_0, \qquad
\tens{S} = J_h^{-1} \tens{S}_0 \tens{F}_h^T.
\end{align}
Using \eqref{app:Q_0}, 
the referential water flux can be
expressed as
\begin{align}
\vec{Q}_w = k J \tens{F}^{-1} \tens{F}^{-T}
\nabla p,
\end{align}
which is consistent with \eqref{eqn:Q_w}.
Similarly, using \eqref{app:S_0}, the
referential stress tensor is given by
\begin{align}
\tens{S} = \tens{S}^e - (\Pi + p) J \tens{F}^{-T},
\end{align}
which is consistent with \eqref{eqn:S},
where we have defined the referential
elastic stress tensor as
$\tens{S}^e = J_h^{-1} \tens{S}_0^e \tens{F}_h^T$. 
A further application of Nanson's formula
shows that the Cauchy stress tensor is given by
\begin{align}
\tens{\sigma} = J_t^{-1} \tens{S}_0
\tens{F}_t^{T} = \tens{\sigma}^e - (\Pi + p) \tens{I},
\end{align}
where the elastic contribution is given by
$\tens{\sigma}^e = J_t^{-1} \tens{S}_0^{e} \tens{F}_t^{T}$. 



\section{Elastic energy and stress of equi-biaxial in-plane deformations}
\label{app:equibiaxial}

We assume the material is subject
to an equi-biaxial deformation in the plane of 
isotropy (i.e.\ the plane of the fibre network).  Thus, 
the total deformation gradient tensor has
the form $\tens{F}_t = \mathrm{diag}(\lambda_\parallel, \lambda_\parallel, \lambda_\perp)$, where $\lambda_\parallel$
and $\lambda_\perp$ represent the in-plane and out-of-plane
stretches, respectively.  The elastic component of the Cauchy
stress tensor will have the representation $\tens{\sigma}^e = \mathrm{diag}(\sigma^e_\parallel, \sigma^e_\parallel, \sigma^e_\perp)$. 

When the principal stretches of the fibre network are equal,
$\lambda_1 = \lambda_2 = \lambda_\parallel$, the strain energy functions associated with the fibres, as given in equations \eqref{Wf} and \eqref{Wtot}, simplify to
\begin{align}
    W_f=\frac{E_f}{2}(\lambda_\parallel-1)^2\quad\text{and}\quad
   W_f=\begin{cases}
   0, & \lambda_\parallel\le1, \\
   \dfrac{E_f(\lambda_\parallel-1)^4(5\lambda_m-3-2\lambda_\parallel)}{2(\lambda_m-1)^3(3+4\lambda_m+3\lambda_m^2)}, & 1\le\lambda_\parallel\le\lambda_m,\\
       \dfrac{E_f}{2}+\dfrac{5E_f\lambda_\parallel(\lambda_\parallel-\lambda_m-1)}{(3+4\lambda_m+3\lambda_m^2)}, & \lambda_\parallel\ge\lambda_m,
   \end{cases} 
\end{align}
respectively. 
The in-plane elastic Cauchy stresses, 
$\sigma^e_{\parallel} = (\lambda_\parallel \lambda_\perp)^{-1} \pdf{W_e}{\lambda_\parallel}$,
are
\begin{align}
\sigma^e_\parallel(\lambda_\parallel,\lambda_\perp)=(1-\Phi_f)\sigma_\text{NH}+\frac{\Phi_fE_f}{2\lambda_\parallel\lambda_\perp}(\lambda_\parallel-1)
    \label{eqn:sigma_para}
\end{align}
and
\begin{align}
\sigma^e_\parallel(\lambda_\parallel,\lambda_\perp)=(1-\Phi_f)\sigma_\text{NH}+\begin{cases}
   0, & \lambda_\parallel\le1, \\
   \dfrac{\Phi_f}{\lambda_\parallel\lambda_\perp} \dfrac{5E_f(\lambda_\parallel-1)^3(1-2\lambda_m+\lambda_\parallel)}{2(\lambda_m-1)^3(3+4\lambda_m+3\lambda_m^2)}, & 1\le\lambda_\parallel\le\lambda_m,\\
       \dfrac{\Phi_f}{\lambda_\parallel\lambda_\perp}\dfrac{5E_f(2\lambda_\parallel-1-\lambda_m)}{2(3+4\lambda_m+3\lambda_m^2)}, & \lambda_\parallel\ge\lambda_m,
   \end{cases} 
\end{align}
respectively, where 
\begin{align}
    \sigma_\text{NH}(\lambda_\parallel,\lambda_\perp)=\frac{E_m}{2(1+\nu_m)}\frac{1}{\lambda_\perp}\left(1-\frac{1}{\lambda_\parallel^2}\right)+\frac{E_m\nu_m}{(1+\nu_m)(1-2\nu_m)}(\lambda_\parallel^2\lambda_\perp-1)
\end{align}
is the stress associated with the neo-Hookean matrix. In both cases, the out-of-plane elastic stress, $\sigma^e_\perp = \lambda_\parallel^{-2} \pdf{W_e}{\lambda_\perp}$, is
\begin{align}
\sigma^e_\perp(\lambda_\parallel,\lambda_\perp)=\frac{(1-\Phi_f)E_m}{1+\nu_m}\left(\frac{1}{2\lambda_\parallel^2}\left(\lambda_\perp-\frac{1}{\lambda_\perp}\right)+\frac{\nu_m}{(1-2\nu_m)}(\lambda_\parallel^2\lambda_\perp-1)\right).
\label{eqn:sigma_perp}
\end{align}

\section{Model reduction for unconfined compression}
\label{app:simplification}

Using the form of the deformation gradient tensor
given by \eqref{eqn:F_cylinder} and assuming that
the referential fluid fraction, radial displacement,
and fluid pressure have the functional forms 
$\phi_w = \phi_w(r,t)$, $u_r = u_r(r,t)$, and $p = p(r, t)$,
the governing equations in Section~\ref{sec:balance_laws} 
can be 
reduced to
\subeq{
  \begin{align}
    \pd{\phi_w}{t} = \frac{1}{r}\pd{}{r}\left[\frac{r k(J) J}{\beta_r^2}\pd{p}{r}\right], 
    \label{app:red_phi_w}
    \\
    \pd{S_r}{r} + \frac{S_r - S_\theta}{r} = 0, 
    \label{app:red_S_r}
    \\
    J = 1 + \phi_w - \phi_0. \label{app:Phi}
  \end{align}}
The boundary conditions for the reduced model are
\subeq{
\begin{align}
  u_r(0,t) = 0, \quad \quad S_r(R,0) = 0. \label{bc:u} \\
  \left.\pd{p}{r}\right|_{r=0} = 0, \quad p(R,t) = 0. \label{bc:p}
\end{align}}
The boundary conditions at $r = 0$
reflect symmetry about the origin.
The conditions at $r = R$ reflect
continuity of total and fluid stress.
In a displacement-controlled experiment, the axial stretch $\beta_z$
is imposed. In a force-controlled experiment, the force
\begin{align}
  F = 2 \pi \int_{0}^{R} (S_z^e - p \beta_r \beta_\theta) r\,\d r
  \label{app:F}
\end{align}
is fixed and $\beta_z$ must be determined from the governing equations.

\subsection{Further reduction of the governing equations}

The above system can be simplified further by noting that
\begin{align}
  \pd{\phi_w}{t} = \pd{}{t}\left(\beta_r \beta_\theta \beta_z\right)
  = \frac{1}{r}\pd{}{r}\left[\frac{r^2}{2} \pd{}{t}\left(\beta_\theta^2 \beta_z\right)\right].
  \label{app:phi_w_t}
\end{align}
Thus, substituting \eqref{app:phi_w_t} into 
the fluid balance equation \eqref{app:red_phi_w},
multiplying by $r$, and integrating in space leads to
\begin{align}
\frac{r}{2}\pd{}{t}\left(\beta_\theta^2 \beta_z\right) = \frac{k(J) J}{\beta_r^2}\pd{p}{r}. \label{app:p}
\end{align}
Moreover, the stress balance \eqref{app:red_S_r}
can be simplified to
\begin{align}
  \pd{S_r^e}{r} + \frac{S_r^e - S_\theta^e}{r} = \beta_\theta \beta_z \left(\pd{p}{r} + \pd{\Pi}{r}\right).
\end{align}
Elimination of the pressure gradient gives
\begin{align}
  \frac{r}{2}\pd{}{t}\left(\beta_\theta^2 \beta_z\right) =
  \frac{k(J)}{\beta_r}\left[  \pd{S_r^e}{r} + \frac{S_r^e - S_\theta^e}{r} - \beta_\theta \beta_z \pd{\Pi}{r}\right],
  \label{app:u}
\end{align}
which is a nonlinear
equation for the radial displacement $u_r$ (which can be seen by
expanding the derivatives of the stretches and elastic stresses). Equation \eqref{app:u}
can be solved with the boundary conditions in \eqref{bc:u}
and the initial conditions $u_r(r,0) = 0$ and $\beta_z(0) = 1$.

\bibliographystyle{IEEEtran}
\bibliography{refs}

\end{document}